\documentclass[12pt]{article}
\usepackage{fullpage}
\usepackage{graphicx}
\usepackage{bbm}
\usepackage{amsmath}
\usepackage{amsfonts,amsbsy}
\usepackage{amssymb}
\usepackage{cite}

\newcommand{\beqa}{\begin{eqnarray}}
\newcommand{\eeqa}{\end{eqnarray}}

\def\x{ {\bf x}}
\def\p{ {\bf p}}
\def\k{ {\bf k}}
\def\q{ {\bf q}}

\def\Ep{ E_\p }
\def\tEp{ {\tilde E}_\p }
\def\Eq{ E_\q }
\def\feq{ f_{\textrm{eq}} }
\newcommand{\n}[1]{n_{{\bf #1}}}
\newcommand{\npr}[1]{n_{{\bf #1}'}}
\def\np{ \n{p} }
\def\nx{ \n{x} }
\def\wp{ \omega_\p }
\def\G{ \overline{\Gamma} }
\def\Gt{ \tilde{\Gamma} }
\def\omegaBar { \frac{\partial \left(\beta\wp\right) }{\partial \beta} }
\def\EBar { \frac{\partial \left(\beta\Ep\right) }{\partial \beta} }
\def\intPSa {\int \frac{d^3\p}{(2\pi)^3} }
\def\intPSb {\int \frac{d^3\p}{(2\pi)^3\Ep} }
\newcommand{\Epa}[2]{ E_{{\bf #1}_{#2}}}
\newcommand{\intPSba}[1]{ \int \frac{d^3\p}{(2\pi)^3\Epa{p}{#1}} }
\def\qbar{ \overline{q} }
\def\as{ \alpha_s }
\def\chiB{\chi_{_\Pi}}
\def\lsim{\mathrel{\rlap{\lower4pt\hbox{\hskip1pt$\sim$}}
    \raise1pt\hbox{$<$}}}
\def\gsim{\mathrel{\rlap{\lower4pt\hbox{\hskip1pt$\sim$}}
    \raise1pt\hbox{$>$}}}

\begin{document}
\title{Bulk viscosity, particle spectra and flow\\ in heavy-ion collisions}
\author{Kevin Dusling and  Thomas Sch\"afer\\[0.2cm]
Department of Physics, North Carolina State University,\\[0.1cm] 
Raleigh, NC 27695.}
\date{\today}
\maketitle

\begin{abstract}
We study the effects of bulk viscosity on $p_T$ spectra and 
elliptic flow in heavy ion collisions. For this purpose we compute 
the dissipative correction $\delta f$ to the single particle distribution 
functions in leading-log QCD, and in several simplified models. We consider, 
in particular, the relaxation time approximation and a kinetic model for
the hadron resonance gas. We implement these distribution functions
in a hydrodynamic simulation of ${\it Au}+{\it Au}$ collisions at 
RHIC. We find significant corrections due to bulk viscosity in hadron 
$p_T$ spectra and the differential elliptic flow parameter $v_2(p_T)$. 
We observe that bulk viscosity scales as the second power of conformality 
breaking, $\zeta\sim \eta(c_s^2-1/3)^2$, whereas $\delta f$ scales as 
the first power. Corrections to the spectra are therefore dominated 
by viscous corrections to the distribution function, and reliable 
bounds on the bulk viscosity require accurate calculations of $\delta f$ 
in the hadronic resonance phase. Based on viscous hydrodynamic simulations 
and a simple kinetic model of the resonance phase which correctly 
extrapolates to the kinetic description of a dilute pion gas  
we conclude that it is difficult to describe the $v_2$ spectra 
at RHIC unless $\zeta/s\lsim 0.05$ near freeze--out.  
We also find that effects of the bulk viscosity on the $p_T$ integrated 
$v_2$ are small. 

\end{abstract}

%%%%%%%%%%%%%%%%%%%%%%%%%%%%%%%%%%%%%%%%%%%%%%%%%%%%%%%%%%%%%%%%%%%%%%%%%
\section{Introduction}
\label{sec_intro}
%%%%%%%%%%%%%%%%%%%%%%%%%%%%%%%%%%%%%%%%%%%%%%%%%%%%%%%%%%%%%%%%%%%%%%%%%

One of the fascinating discoveries of the Relativistic Heavy Ion Collider 
(RHIC) program is the near ideal nature of the fluid produced in the 
collision of two heavy nuclei \cite{Adler:2003kt,Dusling:2007gi,Romatschke:2007mq,Song:2007ux,Schafer:2009dj}.  There is a general consensus in the 
community that the ratio of the shear viscosity to the entropy density 
of the system is no more than a few times the bound $\eta/s \gtrsim 1/4\pi$
conjectured by Kovtun, Son, and Starinets \cite{Kovtun:2004de}.  However, 
it is difficult to determine the level of accuracy that can be obtained 
when extracting the transport properties. To date, the best 
estimate of the shear viscosity comes from a detailed comparison of 
particle spectra and elliptic flow with viscous hydrodynamic simulations 
\cite{Song:2010mg}.  But within these state of the art calculations there 
are many systematic uncertainties which are not fully under control. Some 
of these include the precise form of the initial condition, the details of 
the equation of state, the handling of the freeze--out dynamics, and the 
role of bulk viscosity. Irrespective of its role in constraining shear 
viscosity, the bulk viscosity of the matter produced at RHIC and the LHC 
is clearly an interesting quantity in itself. In this work we will
study the effects of bulk viscosity on the spectra and the elliptic
flow parameter. Our goal is to assess the uncertainty in the extraction
of $\eta/s$ due to the bulk viscosity, and to identify observables that 
constrain the bulk viscosity.

%%%%%%%%%%%%%%%%%%%%%%%%%%%%%%%%%%%%%%%%%%%%%%%%%%%%%%%%%%%%%%%%%%%%%%%%%
\begin{figure}[t]
\begin{center}
\includegraphics[scale=.8]{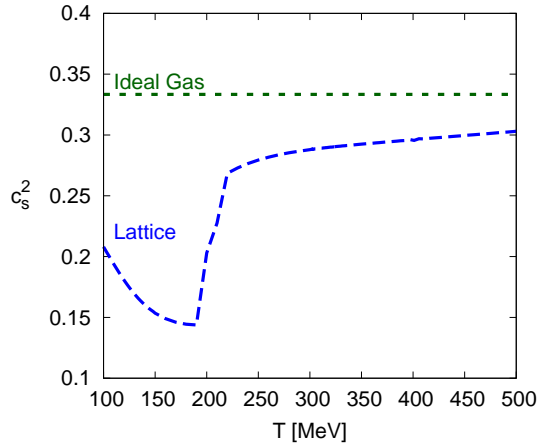}
\caption{Sound speed squared as a function of temperature from the 
parameterization of the lattice QCD equation of state given in 
\cite{Laine:2006cp}. See \cite{Huovinen:2009yb} for a discussion of the 
various parameterizations available for the QCD equation of state.}
\label{fig:cs2}
\end{center}
\end{figure}
%%%%%%%%%%%%%%%%%%%%%%%%%%%%%%%%%%%%%%%%%%%%%%%%%%%%%%%%%%%%%%%%%%%%%%%%%

The earliest viscous hydrodynamic simulations only included corrections 
due to shear viscosity.  One could argue that this may be a safe assumption 
as there are a number of physical systems, possibly relevant to heavy--ion 
collisions, where the bulk viscosity is zero or negligible.  For example, 
it is well known that bulk viscosity vanishes in both the non--relativistic 
and ultra--relativistic limits of a gas when the number of particles are 
conserved \cite{Danielewicz:1984ww}.  In a weakly coupled quark--gluon 
plasma, it was found that the bulk viscosity is on the order of $1000$ 
times smaller than the shear viscosity \cite{Arnold:2006fz}.  Finally, 
in the simplest kinetic model, the relaxation time approximation, 
one finds that the bulk viscosity goes as the square of the deviation 
from conformality,
\beqa
\zeta \approx 15\eta\left(\frac{1}{3}-c_s^2\right)^2\;.
\label{eq:param1}
\eeqa
The above relation was first found by Weinberg for a photon gas coupled 
to matter \cite{Weinberg:1971mx}.  It also happens to give parametrically 
correct results for weakly coupled QCD but not for a scalar field theory.  
In the context of AdS/CFT an analogous relationship \cite{Buchel:2007mf} 
has been found, 
\beqa
\zeta\gtrsim 2\eta\left(\frac{1}{3}-c_s^2\right)\;.
\label{eq:param2}
\eeqa
In this case the bulk viscosity is proportional to the first power of 
conformal breaking.  Based on these above examples, it is clear that for 
a system which is nearly conformally invariant (such as weakly coupled 
QCD) the bulk viscosity will be small.  However, lattice QCD computations 
\cite{Bazavov:2009zn} have shown that the equation of state differs 
strongly from the conformal limit at temperatures relevant to heavy--ion 
collisions (see fig.~\ref{fig:cs2}). For example, if the speed of sound 
approaches $c_s^2\approx 0.2$ near the phase transition we find $\zeta
\approx 0.25\eta$ using either of the expressions (\ref{eq:param1}) 
or~(\ref{eq:param2}) given above.  Even larger values $\zeta \approx 0.6
\eta$ have been obtained in direct lattice studies of the bulk viscosity 
in the regime $T=(1.25-1.65)T_c$ \cite{Meyer:2007dy}. It is therefore 
important to study how bulk viscosity modifies hadronic observables, 
such as $p_T$ spectra and elliptic flow. Previous studies of this type
can be found in \cite{Denicol:2009am,Monnai:2009ad,Song:2009rh,Bozek:2009dw,Bozek:2011ua,Bozek:2011wa,Denicol:2010tr,Roy:2011xt}.

 We begin by reminding the reader how shear viscosity manifests itself 
in the spectra of produced particles.  The equation of hydrodynamics 
express the conservation of the energy momentum tensor, 
\beqa
\partial_\mu T^{\mu\nu}=0\;,
\eeqa
which is given as a sum of ideal and dissipative parts,
\beqa
T^{\mu\nu}=\left(\epsilon+\mathcal{P}\right)u^\mu u^\nu 
 + \mathcal{P}g^{\mu\nu}+\pi^{\mu\nu}+\Pi\Delta^{\mu\nu}\;.
\label{eq:SETdef1}
\eeqa
In the above expression for the stress--energy tensor we have used 
the definition of the three--frame projector $\Delta^{\mu\nu}=g^{\mu\nu}
+u^\mu u^\nu$.  In the first--order (or Navier--Stokes) approximation 
the dissipative parts of the stress--energy tensor can be written in 
the local rest frame as 
\beqa
\pi^{ij}&=& - \eta\left(\partial^i u^j+\partial^j u^i
        -\frac{2}{3}\delta^{ij}\partial_k u^k\right)
  = -\eta\sigma^{ij}\equiv -2\eta\langle \partial^i u^j\rangle\;,\\
\Pi &=&-\zeta \partial_k u^k\;,
\eeqa
where $\eta$ ($\zeta$) is the shear (bulk) viscosity and $\langle \cdots 
\rangle$ indicates that the bracketed tensor should be symmetrized and 
made traceless.  In principle, it would be satisfactory to solve the 
relativistic Navier--Stokes equations in order to compute the first--order 
viscous correction to particle spectra.  However, the first order theory 
is plagued with difficulties such as instabilities and violations of 
causality.  In order to circumvent these difficulties it is necessary 
to use a second order theory, like the one proposed by Israel and 
Stewart \cite{Israel:1976tn,Israel:1979wp} or \"Ottinger and Grmela 
\cite{Grmela:1997zz,Ottinger:1998zz}.  The two theories are qualitatively 
the same in that they both approach the first order theory for small 
relaxation times.  In this work we will not be interested in the 
higher--order corrections arising from the second order theory. Instead 
we use second order hydrodynamics as a practical way to obtain the lowest
order correction in going from ideal to Navier--Stokes hydrodynamics.

\begin{figure}[t]
\begin{center}
\includegraphics[scale=.8]{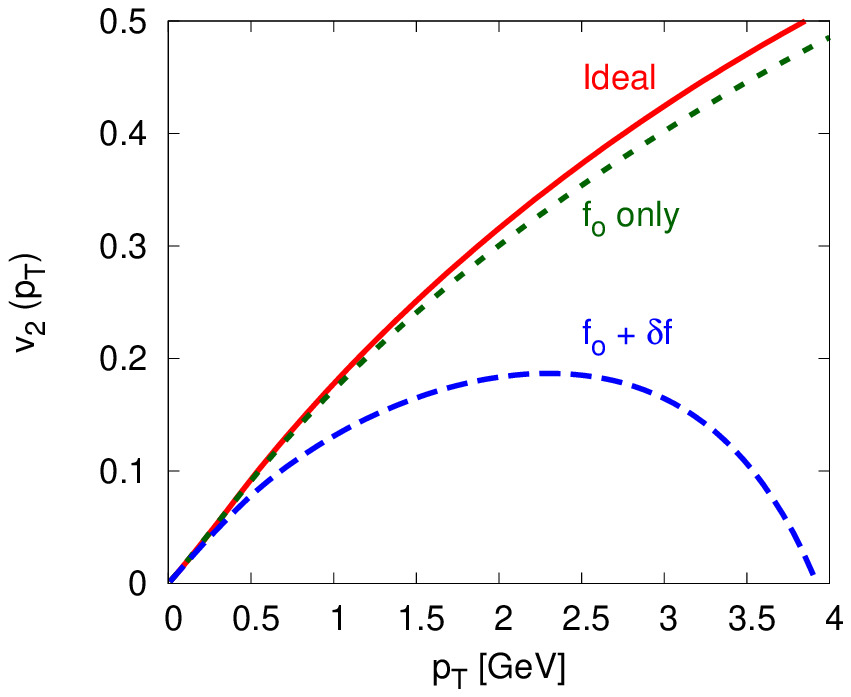}
\includegraphics[scale=.8]{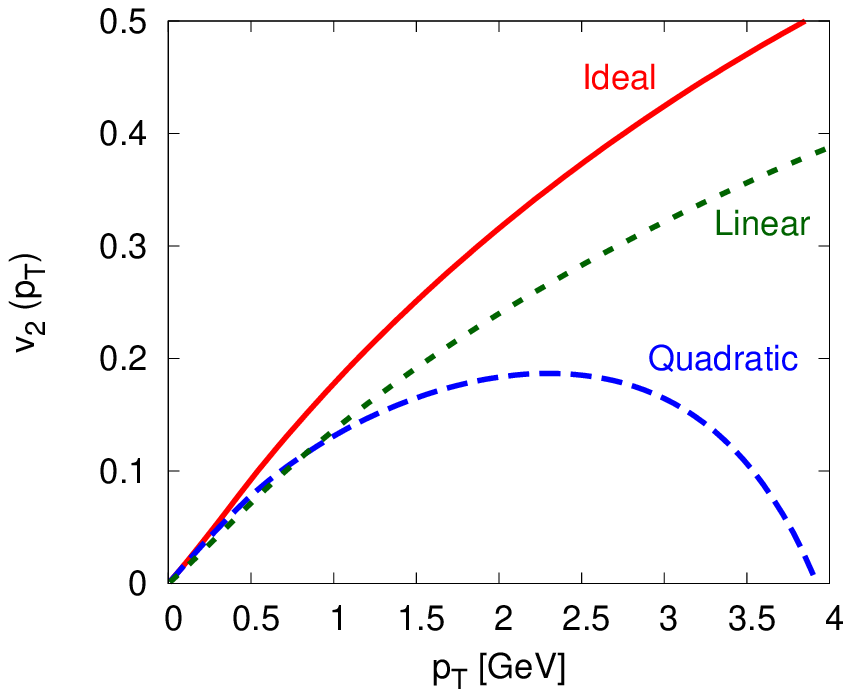}
\caption{Typical results for $v_2(p_T)$ from a viscous hydrodynamic 
model using a quadratic deviation from equilibrium; $\delta f(\p)\sim 
p^2$.  The left plot also shows the viscous result without including 
the off--equilibrium correction to the distribution function.  The 
right plot compares the quadratic ansatz with a linear ansatz; $\delta 
f(\p)\sim p$.  Both curves result in the same shear viscosity to 
entropy ratio.}
\label{fig:quadlin}
\end{center}
\end{figure}

 The solution to the Navier--Stokes equations will lead to viscous 
corrections to the resulting temperature and flow profiles.  Particle 
spectra are then computed using the Cooper--Frye \cite{Cooper:1974mv} 
formula
\beqa
\Ep\frac{dN}{d^3p}=\frac{1}{(2\pi)^3}\int_\sigma f(\Ep) p^\mu d\sigma_\mu\;,
\eeqa
where $\sigma_\mu$ is the freeze--out hypersurface taken as a surface of 
constant energy density in this work.  For a system out of equilibrium 
$f(\Ep)$ is not the equilibrium distribution function but also contains 
viscous corrections
\beqa
f(\Ep) = f_0(\Ep)+\delta f(\Ep)\;,
\eeqa 
where $f_0$ is the usual equilibrium Bose/Fermi distribution function.  
The only constraint on $\delta f$ is that the stress--energy tensor 
remains continuous across the freeze--out hypersurface;
\beqa
\delta T^{\mu\nu}=\intPSb p^\mu p^\nu \delta f(\Ep)\;.
\label{eq:match1}
\eeqa
As shown in \cite{Dusling:2009df} this constraint still leaves a lot 
of freedom in the form of $\delta f$ for shear viscosity.  It was argued that the functional 
form of $\delta f$ could fall anywhere between a linearly increasing 
function of momentum to a quadratically increasing function of momentum.  
These two forms of the distribution function lead to qualitatively 
different behavior for $v_2(p_T)$ as demonstrated by the right plot 
of fig.~\ref{fig:quadlin}. By definition $v_2(p_T)$ is given by
\beqa
v_2(p_T) \equiv \frac{\int d\phi \cos(2\phi)\:( dN +\delta dN)} 
{\int d\phi \: ( dN +\delta dN )}\,,
\label{eq:v2def}
\eeqa
where $dN$ is  short for $dN/({dp_T} \,{d\phi})$ and $\delta  dN$ is 
the first viscous correction to this. If, as a pedagogical exercise 
we neglect the viscous correction to the distribution function all 
together (which violates energy--momentum conservation across the 
freeze--out surface), $v_2(p_T)$ would follow the curve labeled 
`$f_0$' as shown in the left plot of fig.~\ref{fig:quadlin}.  Clearly, 
the form of the viscous correction to the distribution function will 
play an important role in extracting the shear viscosity.  

%%%%%%%%%%%%%%%%%%%%%%%%%%%%%%%%%%%%%%%%%%%%%%%%%%%%%%%%%%%%%%%%%%%%%%%%%
%%%%%%%%%%%%%%%%%%%%%%%%%%%%%%%%%%%%%%%%%%%%%%%%%%%%%%%%%%%%%%%%%%%%%%%%%

There is an analogous viscous correction to the distribution function 
coming from bulk viscosity as well.  The main goal of this work is to 
characterize the functional form of $\delta f$ due to bulk viscosity 
for various theories and models.  We will also show how bulk viscous 
corrections exhibit themselves in spectra as well as some phenomenological 
consequences.

%%%%%%%%%%%%%%%%%%%%%%%%%%%%%%%%%%%%%%%%%%%%%%%%%%%%%%%%%%%%%%%%%%%%%%%%%
\section{The Boltzmann transport equation}
\label{sec_boltzmann}
%%%%%%%%%%%%%%%%%%%%%%%%%%%%%%%%%%%%%%%%%%%%%%%%%%%%%%%%%%%%%%%%%%%%%%%%%

Let us first start by setting up the notation that will be used throughout 
this work. The equilibrium distribution functions for bosons and fermions are
\beqa
\np=\frac{1}{e^{\beta \Ep} \mp 1}\;,
\eeqa
where the upper (minus) sign is for bosons and the lower (plus) sign 
is for fermions.  We will use capital letters $P,Q$ to label 4--vectors 
and bold--type $\p,\q$ for their corresponding 3--vector components 
having energy $\Ep,\Eq$.  The magnitude of the three--momentum will 
be written as $p,q$.  The sign convention for the metric tensor is 
$[-,+,+,+]$ and therefore the hydrodynamic fluid four--velocity obeys 
the normalization condition $u_\mu u^\mu=-1$.  We also use the notation 
$\wp\equiv P_\mu(\beta) u^\mu(t,\x)$ for the quasi--particle's energy 
in the laboratory frame having four momentum $P^\mu=\left(P^0\equiv\Ep,
\p\right)$ in the local rest frame.  

The starting point for our analysis will always be the Boltzmann 
transport equation
\beqa
\mathcal{D}f(t,\x,\p)\equiv  
  \left(\partial_t + v_\p\cdot \partial_\x
            + {\bf F}\cdot\partial_\p \right)f(t,\x,\p)
 = -\mathcal{C}[f,\p]\;,
\label{eq:bltz}
\eeqa
where $v_\p$ is the particle's velocity and ${\bf F}$ is the external 
force on the particle,
\beqa
v_\p\equiv \partial_\p \Ep \;,\;\;\;\;\; 
{\bf F}\equiv \frac{d\p}{dt}=-\partial_\x \Ep\;.
\eeqa
In this work we will consider only small deviations from local thermal 
equilibrium and therefore expand the Boltzmann equation around the
local thermal equilibrium solution
\beqa
\feq(t,\x,\p)=\frac{1}{e^{-\beta(t,\x)\wp(t,\x)}\mp 1}\;.
\eeqa
This procedure is known as the Chapman-Enskog expansion. In the 
Chapman-Enskog procedure we expand the left hand side of the 
Boltzmann equation in gradients of the thermodynamic variables
and linearize the collision operator in $\delta f = f-\feq$. 
Using the following relations\footnote{Two useful identities are 
$\partial \np/\partial p = -\np(1\pm \np)$ and $\partial^2  \np/
\partial p^2=\np(1\pm \np)(1\pm 2\np)$.}
\beqa
\frac{\partial \feq}{\partial \beta} &=& \np\left(1\pm \np\right) 
     \omegaBar \;,\\
\frac{\partial \feq}{\partial \wp} &=& \np\left(1\pm \np\right) \beta \;,
%\frac{\partial \feq}{\partial p^\mu} &=& \np\left(1\pm \np\right) \beta u_\mu\;,
\eeqa
the left--hand side of the Boltzmann equation can be written as\footnote{Even 
though we are working in the local rest frame, gradients that are acting on 
the flow velocity are still non--vanishing.  For example, $\partial_\mu u^i 
\neq 0$ but $\partial_\mu u^0=0$ since $u_\mu u^\mu =-1$.}
\beqa
\label{chap_ens}
\frac{\mathcal{D} \feq}{\np(1\pm \np)} = \EBar \left(\partial_t 
+ v_\p\cdot \partial_\x\right) \beta + \beta \left(\partial_t 
+ v_\p\cdot \partial_\x+{\bf F}\cdot\partial_\p\right) \wp\;.
\eeqa
Let us now assume that the quasi--particles in our system have a dispersion 
relation of the form
\beqa
\Ep=\sqrt{m^2\left(\beta(\x,t)\right)+\p^2}\;,
\label{eq:dispersion}
\eeqa
where we have implicitly included a mass that may be a function of 
temperature.  With this dispersion relation the following identities 
hold
\beqa
v_\p=\frac{{\bf p}}{\Ep}\;,\;\;\;\;\;\;\; {\bf F} 
 = -\frac{m}{\Ep}\partial_{\bf x} m
 = -\frac{\partial \Ep}{\partial \beta} \partial_{\bf x} \beta\;.
\eeqa
Making use of the above relations the left--hand side of the Boltzmann 
equation can be rewritten as\footnote{In deriving this expression 
we have used the two equilibrium identities $\partial_t u_i=\partial_i
\ln\beta$ and $\partial_t\ln\beta=c_s^2\partial_i u^i$.}
\beqa
\frac{\Ep \mathcal{D} \feq}{\beta \np(1\pm \np)} 
 = \frac{1}{2}p^i p^j\sigma_{ij} + \partial_i u^i \left(\frac{p^2}{3}
  - c_s^2 \Ep\frac{\partial \left(\beta\Ep\right) }{\partial \beta}\right)\;,
\label{eq:streamLin}
\eeqa
where we have defined
\beqa
\sigma^{ij}=2 \langle \partial^i u^{j}\rangle
  =\left(\partial^i u^j+\partial^j u^i -
\frac{2}{3}\delta^{ij}\partial_k u^k\right) \;.
\label{eq:evol}
\eeqa
In order to match the kinetic description to hydrodynamics we need to
define a covariantly conserved energy--momentum tensor in the kinetic
theory. There is a subtlety that comes about due to the space--time 
dependence of the mass in the dispersion relation.  In order to see 
this, let us first start with the canonical form of the stress--energy 
tensor which is typically used in kinetic theory
\beqa
T^{\mu\nu}(t,\x)=\intPSb P^\mu P^\nu f(t,\x,\p)\;.
\label{eq:SETdef2}
\eeqa
For situations where the dispersion relation is independent of the medium 
this form is satisfactory as one can show that energy and momentum is 
covariantly conserved\footnote{This can be seen by using the definition 
of the stress--energy tensor given in eq.~(\ref{eq:SETdef2}) and 
differentiating both sides.  For the specific case where the dispersion 
relation is independent of space--time we find 
\beqa
\partial_\mu T^{\mu\nu}=\intPSb p^\nu p^\mu \partial_\mu f(t,\x,\p)\;.
\eeqa
In this case the Boltzmann equation is $p^\mu \partial_\mu  f(t,\x,\p)
=-\Ep \mathcal{C}[f,\p]$ and we find
\beqa
\partial_\mu T^{\mu\nu}=-\intPSa p^\nu \mathcal{C}[f,\p]\;.
\eeqa
The four-momentum is a collisional invariant and the right--hand side vanishes.}
\beqa
\partial_\mu T^{\mu\nu}=0\;.
\eeqa
In the case where we have a non--trivial dispersion relation the partial 
integration can not pass through the integration measure.  Instead we 
find that
\beqa
\partial_\mu T^{\mu\nu}=S^\nu\;,
\eeqa
where
\beqa
S^\nu=\intPSa f(t,\x,\p)\partial_\mu\left(\frac{P^\mu P^\nu}{\Ep}\right)
 - \intPSa P^\nu\; {\bf F}\cdot\partial_\p f(t,\x,\p)\;.
\eeqa
We would like to modify the stress--energy tensor such that the above 
source term vanishes.  This can be achieved by using the definition
\beqa
T^{\mu\nu}=\intPSb \left(P^\mu P^\nu-u^\mu u^\nu T^2
  \frac{\partial m^2}{\partial T^2}\right) f(t,\x,\p)\; .
\label{eq:SETdef3}
\eeqa
Throughout this work we will always use this modified form of the 
stress--energy tensor when matching from the kinetic theory to the 
macroscopic hydrodynamic fields.  We stress that if the quasi--particle's 
mass is space--time independent the above two definitions of the 
stress--energy tensor coincide.  We also note that these observations
are not new.  The modified form of the stress--energy tensor was 
used in studies of the bulk viscosity of a hadronic gas 
\cite{Davis:1991zc,Chakraborty:2010fr,Chen:2011km,Bluhm:2010qf} and of scalar field theory 
\cite{Jeon:1994if,Jeon:1995zm}.

%%%%%%%%%%%%%%%%%%%%%%%%%%%%%%%%%%%%%%%%%%%%%%%%%%%%%%%%%%%%%%%%%%%%%%%%%
\section{Relaxation time approximation}
\label{sec_rta}
%%%%%%%%%%%%%%%%%%%%%%%%%%%%%%%%%%%%%%%%%%%%%%%%%%%%%%%%%%%%%%%%%%%%%%%%%

 In this section we consider the simplest form of the collision kernel, 
which is known as the relaxation time approximation (RTA) or   
Bhatnagar-Gross-Krook (BGK) approximation.  In this model the collision 
term has the simple form
\beqa
\mathcal{C}[f,\p]=\frac{f(\p)-\np}{\tau_R(\Ep)}\;.
\eeqa
If we define the deviation from equilibrium as $\delta f(t,\x,\p)\equiv 
n_p-f(\p)$ and use the linearized form of the streaming operator given 
in eq.~(\ref{eq:streamLin}) we find that
\beqa
\delta f = -\frac{\tau_R(\Ep)}{\Ep T} \np(1\pm \np)
  \left[\frac{1}{2}p^i p^j\sigma_{ij} 
   + \partial_i u^i \left(\frac{p^2}{3}
   - c_s^2 \Ep \frac{\partial \left(\beta\Ep\right) }
     {\partial \beta}\right)\right]\;.
\label{eq:dfRTA}
\eeqa
We would now like to identify the relaxation time encoded in $\delta f$ 
with the transport coefficients $\eta$ and $\zeta$.  First we start with 
the shear viscosity.  Looking at any of the off-diagonal components of 
the stress--energy tensor given in eqs.~(\ref{eq:SETdef1}) and 
(\ref{eq:SETdef3}) we find in the local rest frame
\beqa
\delta T^{xy} = -2\eta\langle \partial^x u^y\rangle 
 = \intPSa \frac{p^x p^y}{\Ep}\delta f\;,
\eeqa
and the shear viscosity can be identified as
\beqa
\eta = \frac{\beta}{30\pi^2}\int \frac{p^6}{\Ep^2}\tau_R(\Ep) 
  \np(1\pm \np)\,dp\;.
\label{eq:etaRTA}
\eeqa
If we take a relaxation time of the form\footnote{We follow the 
notation of \cite{Dusling:2009df} whereby taking $\alpha=0$ corresponds 
to the usual quadratic ansatz.  In this case the relaxation time grows 
linearly with momentum, $\tau_R\sim \Ep$, and $\chi\sim p^2$.  The other 
extreme case follows from $\alpha=1$ where now the relaxation is independent 
of momentum, $\tau_R\sim \rm{Const.}$, and $\chi\sim p$.  For leading 
order QCD one numerically finds $\alpha=0.62$ and $\chi\sim p^{1.38}$. }
\beqa
\tau_R(\Ep)=\tau_0\beta\left(\beta \Ep\right)^{1-\alpha}\;,
\eeqa
we find the following relation between the shear viscosity and relaxation 
time
\beqa
\eta=\frac{\tau_0 T^3}{30\pi^2} \mathcal{I}_\alpha(\beta m) \;,
\label{eq:etaRTA2}
\eeqa
where the dimensionless phase space integral $\mathcal{I}_\alpha$ is worked 
out in appendix~\ref{app:RTA}.

We now come to bulk viscosity, which characterizes the deviation of the 
pressure from its equilibrium value as the fluid expands or contracts more 
quickly than the time it takes the pressure to relax back to its equilibrium 
value. The bulk viscous pressure, $\Pi$, is therefore related to the extra 
pressure from the departure from equilibrium $\delta f$.  However, 
the departure from equilibrium can not only shift the pressure but 
also the energy density by an amount $\delta \epsilon$.  This shift in 
energy density will also lead to a shift in pressure, which should 
not be included in the bulk viscous pressure.  This is because the 
bulk viscous pressure should only include the difference between the 
actual pressure and the pressure determined by thermodynamics 
\cite{Arnold:2006fz} which in our case will be $\mathcal{P}(\epsilon 
+ \delta \epsilon)$.  This additional pressure shift must therefore 
be subtracted when defining the bulk viscous pressure\footnote{We have 
used 
\beqa
\mathcal{P}(\epsilon_0+\delta\epsilon)
  \approx \mathcal{P}(\epsilon_0)+c_s^2\delta e\;,
\label{eq:BP1}
\eeqa
where from eq.~(\ref{eq:SETdef3}) we have
\beqa
\delta\epsilon = \intPSb\left(\Ep^2
  - T^2\frac{\partial m^2}{\partial T^2}\right)\delta f\;.
\label{eq:BP2}
\eeqa},
\beqa
\Pi \equiv \frac{1}{3}T^{ii} - \mathcal{P}(\epsilon + \delta \epsilon)
 = \intPSb \left(\frac{p^2}{3}
   - c_s^2 \Ep \frac{\partial \left(\beta\Ep\right) }
                       {\partial \beta}\right)\delta f\;.
\label{eq:bulkdf}
\eeqa
Making use of the form of the dispersion relation in eq.~(\ref{eq:dispersion}) 
it will be convenient to define the quantities $\tilde{m}$ and $\tEp$ 
via
\beqa
\Ep\frac{\partial \left(\beta\Ep\right) }{\partial \beta}
 = p^2+\left(m^2-\frac{\partial m^2}{\partial T^2}T^2\right)
 \equiv p^2+\tilde{m}^2\equiv \tEp^2\;.
\eeqa
The following relation between the relaxation time and bulk viscosity 
coefficient $\zeta$ then holds,
\beqa
\zeta = \frac{\tau_0 T^3}{2\pi^2} 
  \mathcal{J}_\alpha(\beta m,\beta \tilde{m})\;,
\eeqa
where the dimensionless phase space integral $\mathcal{J}_\alpha$ depends 
on both the thermal mass $m$ and the shifted mass $\tilde{m}$.  This phase 
space integral is discussed at length in appendix~\ref{app:RTA}. In the high 
temperature limit, $(T\gg m,\tilde{m})$, one finds 
\beqa
\eta = \frac{\tau_0 T^3}{30\pi^2}\G(6-\alpha)\;,\;\;\;\;\;\; 
\zeta=\frac{\tau_0 T^3}{2\pi^2}\G(6-\alpha)
              \left(\frac{1}{3}-c_s^2\right)^2 \;,
\eeqa
where the function $\G$, defined in appendix~\ref{app:RTA}, depends on the 
statistics of the particles.  For classical statistics $\G$ is the usual 
Gamma function. From the above formulas we can recover the well--known 
relationship \cite{Weinberg:2008zzc} between shear and bulk viscosity,
\beqa
\label{zeta_scal_rta}
\zeta=15\eta\left(\frac{1}{3}-c_s^2\right)^2\;.
\eeqa
We note that this relation is independent of the momentum dependence of 
the relaxation time.

%%%%%%%%%%%%%%%%%%%%%%%%%%%%%%%%%%%%%%%%%%%%%%%%%%%%%%%%%%%%%%%%%%%%%%%%%
\subsection{Landau matching in the relaxation time approximation}
\label{sec_match}
%%%%%%%%%%%%%%%%%%%%%%%%%%%%%%%%%%%%%%%%%%%%%%%%%%%%%%%%%%%%%%%%%%%%%%%%%

Landau matching is a way to uniquely specify the energy density $\epsilon$ 
and fluid four velocity $u^\mu$ in terms of four components of $T^{\mu\nu}$.  
If we use the Landau--Lifshitz convention
\beqa
\epsilon &=& u_\mu u_\nu T^{\mu\nu}\;,\\
\epsilon u^\mu &=& -u_\nu T^{\mu\nu}\;,
\eeqa
then the other six independent components of $T^{\mu\nu}$ are given 
by a non-equilibrium stress tensor $\pi^{\mu\nu}$ satisfying $u_\mu 
\pi^{\mu\nu}=0$. In order that the stress--energy tensor remains 
continuous across the freeze--out surface the functional form of 
$\delta f$ must be such that the Landau matching condition is satisfied; 
$u_\mu \delta T^{\mu\nu}=0$.  From eq.~(\ref{eq:SETdef3}) the matching 
condition is 
\beqa
0 = \intPSb\left(\wp P^\nu + u^\nu T^2\frac{\partial m^2}{\partial T^2}\right) 
  \delta f(\Ep)\;.
\eeqa
It is sufficient for the above matching condition to be satisfied in 
the local rest frame. This corresponds to the condition that the shift 
in energy density stemming from $\delta f$ vanishes, 
\beqa
\delta \epsilon =0=\intPSb \tEp^2\; \delta f(\Ep)\;.
\eeqa
Let us now look at the energy density shift coming from the off--equilibrium 
distribution given in eq.~(\ref{eq:dfRTA})
\beqa
\delta \epsilon_{\textrm{RTA}} 
 = \frac{\Pi\beta^5}{\mathcal{J}_\alpha(\beta m, \beta\tilde{m})} 
    \intPSa\left(\frac{\tEp}{\Ep}\right)^2\np(1\pm \np)
     \left(\frac{p^2}{3}-c_s^2 \tEp^2\right)
     \left(\beta\Ep\right)^{1-\alpha} \;.
\label{eq:deRTA}
\eeqa
The above expression simplifies considerably when there are no mean fields, 
$\tEp\to \Ep$,
\beqa
\delta \epsilon_{\textrm{RTA}} 
  \propto \intPSa\np(1\pm \np)\left(\frac{p^2}{3}-c_s^2 \Ep^2\right)
       \left(\beta\Ep\right)^{1-\alpha} \;.
\eeqa
The above integral vanishes only for $\alpha=1$, which is the case where 
the relaxation time $\tau_R(\Ep)$ is momentum--independent\footnote{This 
is easily seen by using the definition of the sound speed,
\beqa
c_s^2 = \frac{\frac{1}{3}\intPSa p^2\np(1\pm \np)}{\;\;\;
  \intPSa \tEp^2\np(1\pm \np)}\;.
\eeqa
}. Therefore, if one considers a gas of particles where the deviation 
from conformality comes from the bare mass of the particle only (no mean 
fields), then the relaxation time approximation can be used if and only 
if the relaxation time is independent of momentum.  

 In the presence of mean--fields ({\em i.e.} the quasi--particle's mass 
is temperature dependent) we can write eq.~(\ref{eq:deRTA}) as
\beqa
\delta \epsilon_{\textrm{RTA}} &\propto& \intPSa \np(1\pm \np)
     \left(\frac{p^2}{3}-c_s^2 \tEp^2\right)
     \left(\beta\Ep\right)^{1-\alpha} \nonumber\\
 & & \mbox{} - \frac{\partial m^2}{\partial T^2}\intPSa \np(1\pm \np)
      \left(\frac{p^2}{3}-c_s^2 \tEp^2\right)
      \left(\beta\Ep\right)^{-\alpha-1}\;.
\eeqa
In this case taking $\alpha=1$ makes the first term vanish, but the 
second term remains finite (even though it may be parametrically small 
since it is proportional to the coupling). It is possible, however, 
to use the relaxation time approximation consistent with Landau 
matching by a fine--tuning of the parameter $\alpha$.

%%%%%%%%%%%%%%%%%%%%%%%%%%%%%%%%%%%%%%%%%%%%%%%%%%%%%%%%%%%%%%%%%%%%%%%%%
\section{Scalar field theory}
\label{sec_phi4}
%%%%%%%%%%%%%%%%%%%%%%%%%%%%%%%%%%%%%%%%%%%%%%%%%%%%%%%%%%%%%%%%%%%%%%%%%

The case of a weakly coupled scalar field theory was studied  by Jeon 
\cite{Jeon:1994if} where the Boltzmann equation and collision kernel 
were derived from first principles. While the full computation of the 
transport coefficients are numerically intensive a lot can be said 
about the form of the off--equilibrium distribution function from 
certain general considerations. As shown in \cite{Jeon:1995zm} one 
can compute the transport coefficients in $g\phi^3+\lambda\phi^4$ 
theory at weak coupling by solving Boltzmann equation\footnote{For 
our discussion it will be sufficient to look at a pure $\lambda\phi^4$ 
theory.},
\beqa
\left(\partial_t + v_\p\cdot \partial_\x
     + {\bf F}\cdot\partial_\p \right)f(t,\x,\p)
= - \mathcal{C}_{2\leftrightarrow 2}[f,\p]
  - \mathcal{C}_{2\leftrightarrow 4}[f,\p]\;,
\eeqa
where the collision operator has been split into a term containing 
$2\leftrightarrow 2$ processes and a second term involving number 
changing $2\leftrightarrow 4$ processes.  While the number changing 
processes are higher order in the coupling constant ($\lambda$), they 
are required in order for a system undergoing a uniform expansion or 
contraction to equilibrate.  If number changing processes were not 
included the above Boltzmann equation would have no solution. Formally, 
this is due to the presence of a (spurious) zero mode associated with 
particle number conservation in the $2\leftrightarrow 2$ processes.  
This zero--mode is not orthogonal to the source term and subsequently 
renders the linearized Boltzmann equation non--invertible.  We should 
also point out that there is a zero mode corresponding to energy 
conservation.  This zero--mode is not problematic since it is orthogonal 
to the source.  

It is precisely the above behavior of a scalar field theory that allows 
one to obtain the approximate form of the off--equilibrium distribution 
function.  In order to see how this works out let us start by linearizing 
the above Boltzmann equation around its equilibrium solution
\beqa
\label{eq:del_f}
\delta f(\p) = -\np(1+\np)\chi_\pi(p)\hat{p}^i\hat{p}^j\langle 
      \partial_i u_j\rangle - \np(1+\np)\chiB(p)\partial_k u^k\;.
\eeqa
This equation for $\delta f$ follows from the Chapman-Enskog 
expansion eq.~\ref{chap_ens}. 
The equations in the shear and bulk channels can be separated.  
In the spin 0 (bulk) channel we find
\beqa
\frac{\beta}{\Ep}\left(\frac{p^2}{3} 
  - c_s^2 \Ep \frac{\partial \left(\beta\Ep\right) }{\partial \beta}\right)
= -\mathcal{C}_{2\leftrightarrow 2}[\delta f,\p]
  -\mathcal{C}_{2\leftrightarrow 4}[\delta f, \p]\;,
\label{eq:spin0}
\eeqa
where we have written $\mathcal{C}[\delta f,\p]$ to make it explicit that 
the collision term should be linearized around the equilibrium solution.  
The resulting operators (including the final state symmetry factors) are
\beqa
C_{2\leftrightarrow 2}[\delta f,\p] 
  &=& \frac{1}{2!}\int_{\k,\p',\k'} 
  \Gamma_{\p \k\rightarrow \p'\k'}\;\;
      \np\n{k}(1+\npr{p})(1+\npr{k})\nonumber\\
&&\;\;\;\;\;\;\;
   \times\left[ \chiB(p) + \chiB(k)-\chiB(p')-\chiB(k')\right]\;,\\
C_{2\leftrightarrow 4}[\delta f,\p]
  &=& \frac{1}{3!2!}\int_{\k,\p',\k',\q,\q'} 
  \Gamma_{\p' \k\rightarrow \p\k'\q\q'}\;\;
       \np\npr{k}\n{q}\npr{q}(1+\npr{p})(1+\n{k})\nonumber\\
&&\;\;\;\;\;\;\;\;\;\;\;\;\;
  \times\left[ \chiB(p') + \chiB(k)-\chiB(p)
          -\chiB(k')-\chiB(q)-\chiB(q')\right]\nonumber\\
&-& \frac{1}{4!1!}\int_{\k,\p',\k',\q,\q'} 
   \Gamma_{\p \k\rightarrow \p'\k'\q\q'}\;\;
     \npr{p}\npr{k}\n{q}\npr{q}(1+\np)(1+\n{k})\nonumber\\
&&\;\;\;\;\;\;\;\;\;\;\;\;\;
 \times\left[ \chiB(p) + \chiB(k)-\chiB(p')-\chiB(k')-\chiB(q)-\chiB(q')\right]
\eeqa
where we have used the shorthand $\int_\p=\int\frac{d^3\p}{(2\pi)^3}$. The 
transition rates are given as
\beqa
\label{eq:Gamma}
\Gamma_{\p\k\rightarrow \p'\k'}
 = \frac{\left| \mathcal M_{2\rightarrow 2} \right|^2 }{(2 E_\p) 
  (2 E_\k) (2 E_{\p'}) (2E_{\k'}) }  (2\pi)^4
   \delta^4(P + K - P' - K') \, ,
\eeqa
\beqa
\label{eq:Gamma23}
\Gamma_{\p\k\rightarrow \p'\k'\q\q'}
 = \frac{\left| \mathcal M_{2\rightarrow 4} \right|^2 }{(2 E_\p) 
  (2 E_\k) (2 E_{\p'}) (2E_{\k'}) (2E_{\q}) (2E_{\q'}) }  (2\pi)^4
   \delta^4(P + K - P' - K'- Q - Q').
\eeqa
Formally, we can solve eq.~\ref{eq:spin0} by inverting the collision 
operator. Lu and Moore observed that the largest contribution will 
come from the near--zero mode \cite{Lu:2011df} which has the form
\beqa
\label{chiB_phi4}
\chiB(p) =\chi_0-\chi_1 E_p\;,
\eeqa
where $\chi_i$ are constants to be determined.  Substituting the above 
form of $\chiB(p)$ into the spin 0 channel of the linearized Boltzmann 
equation, eq.~(\ref{eq:spin0}), and integrating both sides over all phase 
space we obtain
\beqa
\chi_0=\frac{\beta\mathcal{F}}{4\Gamma_{\text{inelastic}}}\;,
\eeqa
where
\beqa
\Gamma_{\textrm{inelastic}}=\frac{1}{48}\int_{\p\k\p'\k'\q\q'}
  \Gamma_{\p \k\rightarrow  \p'\k'\q\q'}\;\;
  \npr{p}\npr{k}\n{q}\npr{q}(1+\n{p})(1+\n{k})\;,
\label{eq:inTot}
\eeqa
and we have defined the function
\beqa
\mathcal{F} \equiv \intPSb \left(\frac{p^2}{3}
  - c_s^2 \Ep \frac{\partial \left(\beta\Ep\right) }{\partial \beta}\right)
    \np(1+\np)\;,
\label{eq:F}
\eeqa
which characterizes the deviation of the theory from conformality.  The 
total inelastic cross--section given in eq.~(\ref{eq:inTot}) can be 
computed by doing the phase space integrals numerically. From a 
phenomenological perspective this is not necessary.  Instead, the 
total inelastic cross--section can be related to the bulk viscosity 
coefficient by using eq.~(\ref{eq:bulkdf}). This identification leads to
\beqa
\chi_0=\frac{\zeta}{\mathcal{F}}\;.
\eeqa
The constant $\chi_1$ is undetermined by the Boltzmann equation. Instead 
it is constrained by requiring that the deviation from equilibrium does 
not bring about a shift in the energy density, 
\beqa
\delta \epsilon = 0 = \intPSb \tEp^2 \delta f\;.
\eeqa
We therefore find the following form for the off--equilibrium distribution 
function
\beqa
\chiB(p)=\frac{\zeta}{\mathcal{F}}\left(1-\mathcal{G}\Ep\right) \;,
\eeqa
where $\mathcal{F}$ has been defined in eq.~(\ref{eq:F}) and 
\beqa
\mathcal{G}\equiv \frac{\intPSb \tEp^2\np(1+\np)}{\intPSa \tEp^2\np(1+\np)} \;.
\eeqa
For completeness, it is worth discussing the parametric behavior of the 
bulk viscosity at high temperature.  The bulk viscosity coefficient is 
given by
\beqa
\zeta=\frac{\beta\mathcal{F}^2}{4\Gamma_{\textrm{inelastic}}} \;.
\eeqa
In the high temperature limit we can evaluate $\mathcal{F}$ 
semi--analytically (see appendix~\ref{app:scalar}). In this regime we 
can ignore the bare and thermal mass of the scalar quasi--particles (up 
to logarithms). The deviation from conformality contained in $\mathcal{F}$ 
is controlled by the running of the coupling.  For a scalar field theory 
we have
\beqa
m_{\textrm{thermal}}^2 = \frac{\lambda T^2}{24}\longrightarrow \tilde{m}^2
  = \frac{\beta(\lambda)T^2}{48}\;.
\eeqa
and using $\beta(\lambda)=\frac{3\lambda^2}{16\pi^2}$ we find that
\beqa
\mathcal{F} = \frac{\lambda^2 T^4\ln\left(\gamma\lambda\right)}{3(32\pi^2)^2}
  \;\;\;\;\textrm{where}\;\;\;\; 
\gamma\equiv \frac{1}{96}e^{15\zeta_+(3)/\pi^2}\;.
\eeqa
Naively the total inelastic rate would go as $\lambda^4T^4$.  However, 
there is a soft enhancement  which leads to $\Gamma =\# \lambda^3T^4$ 
\cite{Jeon:1995zm}. We therefore find that
\beqa
\zeta = \frac{\lambda^3 T^3\ln^2\left(\gamma\lambda\right)}
  {\#9(32\pi^2)^4}\; .
\eeqa

%%%%%%%%%%%%%%%%%%%%%%%%%%%%%%%%%%%%%%%%%%%%%%%%%%%%%%%%%%%%%%%%%%%%%%%%%
\section{Leading log treatment in QCD}
\label{sec_lla}
%%%%%%%%%%%%%%%%%%%%%%%%%%%%%%%%%%%%%%%%%%%%%%%%%%%%%%%%%%%%%%%%%%%%%%%%%

In this section we will use the Boltzmann equation in the leading 
$\log(T/m_D)$ approximation. In this approximation the dynamics can 
be summarized by a Fokker--Plank equation which describes the momentum 
diffusion of the quasi--particles. The functional form of $\chiB$ can 
be found by solving a simple ordinary differential equation.  We start 
by discussing the pure glue theory and then consider a multi--component 
QGP.   

%%%%%%%%%%%%%%%%%%%%%%%%%%%%%%%%%%%%%%%%%%%%%%%%%%%%%%%%%%%%%%%%%%%%%%%%%
\subsection{Pure Glue}
\label{sec_glue}
%%%%%%%%%%%%%%%%%%%%%%%%%%%%%%%%%%%%%%%%%%%%%%%%%%%%%%%%%%%%%%%%%%%%%%%%%

\begin{figure}[t]
\begin{center}
\includegraphics[scale=.8]{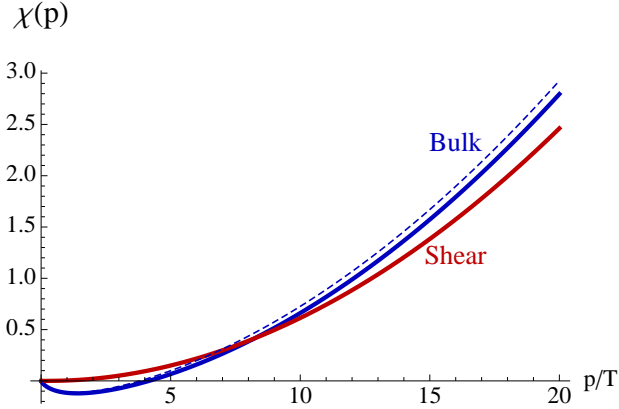}
\caption{Non-equilibrium distribution functions $\chi_\pi$ (red curve
labeled shear) and $\chiB$ (blue curves labeled bulk) of gluons in 
leading log approximation. The functions $\chi_\pi$ and $\chiB$ are 
defined in eq.~(\ref{eq:del_f}). We have rescaled $\chi_\pi$ by one 
power of the conformal breaking parameter, $(1/3-c_s^2)$,
in order to check the expected scaling behavior $\chiB\sim (1/3-c_s^2)
\chi_\pi$. The dotted line shows the bulk viscous correction $\chiB$ 
before it was made orthogonal to the energy density. The curves in
this plot were obtained for $m_D/T=1$, corresponding to a very weak
coupling $\alpha_s=1/(4\pi)$. }
\label{fig:ChiPureGlue}
\end{center}
\end{figure}

In a leading log approximation, $\log(T/m_D)$ is considered to be 
parametrically large.  The resulting dynamics describes Coulomb 
scattering with a small momentum transfer of order $q\sim gT$ but 
with a rapid collision rate of $\sim g^2T$ (up to logarithms).  At 
leading log order the linearized Boltzmann equation can be recast 
as a Fokker-Planck equation \cite{Hong:2010at,Arnold:2000dr}. This 
equation allows us to determine $\chi(p)$ in a suitable limit 
(absence of ``gain'' terms) by solving a differential equation
rather than an integral equation. The Fokker-Planck equation is
\beqa
\frac{1}{2}p^i p^j\sigma_{ij} + \partial_i u^i \left(\frac{p^2}{3}
   -c_s^2 \Ep \frac{\partial \left(\beta\Ep\right)} {\partial \beta}\right)
 &=&  \frac{T\mu_A}{\np(1+\np)}\frac{\partial}{\partial \p^i}
      \left(\np(1+\np)\frac{\partial}{\partial \p^i} 
      \left[\frac{\delta f(\p)}{\np(1+\np)}\right]\right)\nonumber\\
  & & \mbox{} + \frac{\textrm{gain terms}}{\np(1+\np)} \;,
\label{eq:FPglue}
\eeqa
where $\mu_A$ is the drag coefficient in the leading log approximation
\beqa
\frac{d\p}{dt}=\mu_A \hat{\p}\;,\;\;\;\;\;\; 
  \mu_A = \frac{g^2 C_A m_D^2}{8\pi}\log\left(\frac{T}{m_D}\right) \;.
\eeqa
The Debye mass is given by $m_D^2=\frac{1}{3}(C_A+\frac{N_f}{2})g^2T^2$
with $C_A=N_c$. Eq.~(\ref{eq:FPglue}) without the gain terms is a 
Fokker--Planck equation for a hard particle undergoing drag and diffusion 
in a thermal bath. In order to conserve energy and momentum the gain terms 
must be included. The gain terms can be written as \cite{Hong:2010at}
\beqa
\textrm{gain terms}\equiv 
 \frac{6}{T^3}\left[\frac{1}{p^2}\frac{\partial}{\partial p} 
      p^2\np(1+\np)\right]\frac{dE}{dt}
+\frac{6}{T^3}\left[\frac{\partial }{\partial \p} 
    \np(1+\np)\right]\cdot \frac{d {\bf P}}{dt} \;,
\eeqa
where $dE/dt$ and $d{\bf P}/dt$ are the energy and momentum transfer to 
the hard particle from the thermal bath per unit time;
\beqa
\frac{dE}{dt} = \intPSa \hat{\p}\cdot {\bf j}_p\;,\;\;\;\;\;\; 
\frac{d{\bf P}}{dt}=\intPSa {\bf j}_p\;,
\eeqa
where
\beqa
{\bf j}_p = -T\mu_A\np(1+\np)\frac{\partial}{\partial \p}
   \left[\frac{\delta f}{\np(1+\np)}\right]\;.
\eeqa
We express the off--equilibrium distribution function in terms of $\chi_\pi$ 
and $\chiB$ as in equ.~(\ref{eq:del_f}). Substituting this expression into 
the Fokker--Planck equation we find that the shear and bulk contributions 
decouple.  In the shear shear channel the gain terms vanish and we are left 
with the following ordinary differential equation for $\chi_\pi(p)$
\beqa
\frac{p}{T}=\mu_A T\left(-\chi_\pi^{\prime\prime}+\left(\frac{1+2\np}{T}
  - \frac{2}{p}\right)\chi_\pi^{\prime} + \frac{6}{p^2}\chi_\pi\right)\;.
\eeqa
At high momentum $(1+2\np)\to 1$ and we find \cite{Dusling:2009df}
\beqa
\chi_\pi(p) = \frac{1}{2T\mu_A} p^2\;.
\eeqa
The above differential equation can also be solved numerically. For this
purpose two boundary conditions must be specified. The first boundary 
condition is that $\chi_\pi(p=0)=0$, which implies that in QCD soft gluons
equilibrate rapidly. The second boundary condition follows from the 
structure of the solution at large momentum. In general the differential 
equation has two independent solutions; one being a polynomial in $p$ 
and the other growing exponentially in $p$. We choose the second boundary 
condition so that the exponentially growing solution is suppressed. In 
practice, this can be done using a shooting method on $\chi^\prime(p=0)$ 
such that $\chi^{\prime\prime\prime}(p=p_{\rm{max}})=0$, which removes 
the exponential solution.  The result of this procedure is shown in 
fig.~\ref{fig:ChiPureGlue}. The shear viscosity can be found using 
the relation 
\beqa
\label{eta_ll}
\eta=\sum_a \frac{\nu_a}{30\pi^2}\int \frac{p^4}{\Ep} 
   \np(1\pm \np)\chi_\pi(p)\;,
\eeqa
and we find $\eta/\left(g^4 T^3 \ln\right)=27.1$ in agreement with 
\cite{Arnold:2000dr}.

%%%%%%%%%%%%%%%%%%%%%%%%%%%%%%%%%%%%%%%%%%%%%%%%%%%%%%%%%%%%%%%%%%%%%%%%%
%%%%%%%%%%%%%%%%%%%%%%%%%%%%%%%%%%%%%%%%%%%%%%%%%%%%%%%%%%%%%%%%%%%%%%%%%

In the case of bulk $(l=0)$ channel, while $d{\bf P}/dt$ is zero the 
gain term $dE/dt$ is non--vanishing. In order to understand the role 
of this term we first analyze the Fokker--Planck without the gain
term
\beqa
\left(\frac{1}{3}-c_s^2\right)\frac{p}{T}-c_s^2\tilde{m}_A^2\frac{1}{p T}
 = \mu_A T\left(-\chiB^{\prime\prime}
  + \left(\frac{1+2\np}{T}-\frac{2}{p}\right)\chiB^{\prime} \right).
\label{eq:bulkODE}
\eeqa
This differential equation has one exact zero mode, $\chiB\propto {\rm 
const.}$, related to particle number conservation in  $2\leftrightarrow 2$
scattering. This zero mode is removed if $2\leftrightarrow 3$ splitting 
and joining processes are included. We can take this into account by 
imposing the boundary condition $\chiB(p=0)=0$. The second boundary 
condition is chosen in order to suppress the exponentially growing 
solution as discussed in the shear case.

 The solution obtained in this way is not physically acceptable 
because it does not respect energy conservation. The fact that the 
collision term conserves energy implies that the most general solution 
of the linearized Boltzmann equation must be of the form $\chiB(p)=
\chiB^0(p)+\chi^1 p$, where $\chi^1$ is a constant and we have used the fact 
that the leading log collision integral is computed using $E_p\simeq p$. 
It is easy to see that this is a property of the Fokker--Planck equation 
in the bulk channel with the gain term included, but not without it. We 
find that restoring the zero mode $\chiB\propto p$ is the dominant 
effect of the gain term, and that $\chiB^0(p)$ is very well approximated 
by the solution of the ordinary differential equation (\ref{eq:bulkODE}).
 
%%%%%%%%%%%%%%%%%%%%%%%%%%%%%%%%%%%%%%%%%%%%%%%%%%%%%%%%%%%%%%%%%%%%%%%%%
\begin{figure}[t]
\begin{center}
\includegraphics[scale=.8]{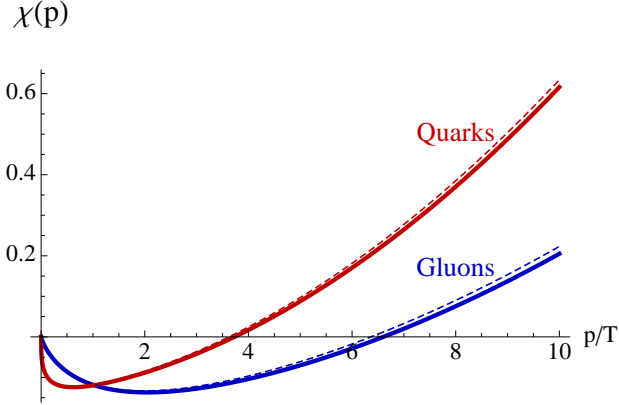}
\caption{Deviation of Quarks and Gluons from equilibrium due to a 
bulk stress in leading log approximation.  The dashed curves show the 
results before the solutions were made orthogonal to the energy density. }
\label{fig:LL}
\end{center}
\end{figure}
%%%%%%%%%%%%%%%%%%%%%%%%%%%%%%%%%%%%%%%%%%%%%%%%%%%%%%%%%%%%%%%%%%%%%%%%%

 The freedom in adding the zero mode has no effect on the calculation 
of the bulk viscous pressure via eq.~(\ref{eq:bulkdf}), because any 
shift in the pressure due to a shift in the energy density is projected 
out. However, in this work we are also interested in the correction to
the single particle spectra, and in that context the linear term in 
$\chiB$ matters. We therefore fix $\chi^1$ by the requirement that 
$\delta f$ does not contribute to the energy density as required by 
the Landau matching conditions
\beqa
0=\intPSb \tEp^2 \np(1+\np)\left[\chiB(p)-\chi^1 p\right]\;.
\eeqa
In this case there is no need to remove the shift in pressure due to 
the shift in energy density when computing the bulk viscosity
\beqa
\label{zeta_ll}
\zeta = \intPSb \frac{p^2}{3} \np(1+\np)\left[\chiB(p)-\chi^1 p\right]\;.
\eeqa
The numerical solution of eq.~(\ref{eq:bulkODE}) is shown in 
fig.~(\ref{fig:ChiPureGlue}). We observe that $\chi_B$ changes sign 
at $p\sim 4T$, and that for large values of the momentum, $p\gsim 
7T$, the non-equilibrium distribution function in the bulk channels 
scales as the distribution function in the shear channel multiplied
by one power of the conformal symmetry breaking parameter 
\beqa 
\label{chib_scal}
\chiB \sim \left(\frac{1}{3}-c_s^2\right) \chi_\pi\, . 
\eeqa
Integrating the solution gives $\zeta/(T^3\alpha_s^2)\ln = 0.44$, in 
agreement with the result in \cite{Arnold:2006fz}. The bulk viscosity 
scales as the second power of the conformal symmetry parameter, 
\beqa 
\label{zeta_scal}
\zeta \sim 47.9 \left(\frac{1}{3}-c_s^2\right)^2 \eta \, . 
\eeqa
This result has the same structure as the relation obtained in 
the relaxation time approximation, eq.~(\ref{zeta_scal_rta}), 
but with a larger numerical coefficient. 

%%%%%%%%%%%%%%%%%%%%%%%%%%%%%%%%%%%%%%%%%%%%%%%%%%%%%%%%%%%%%%%%%%%%%%%%%
\subsection{Quark--Gluon Plasma}
\label{sec_qgp}
%%%%%%%%%%%%%%%%%%%%%%%%%%%%%%%%%%%%%%%%%%%%%%%%%%%%%%%%%%%%%%%%%%%%%%%%%

The previous analysis can be easily extended to a multi--component system.  
For a quark--gluon plasma the extension of eq.~(\ref{eq:bulkODE}) is 
\cite{Hong:2010at,Arnold:2000dr}
\beqa
q_A(p) &=& \mathcal{C}_\textrm{Loss}(\chi^g) 
    - \frac{2\gamma}{p} \frac{N_f d_F}{d_A}\frac{\np^F}{\np^B}
     \left(\chi^q + \chi^{\qbar} - 2\chi^g \right) \;,\\ 
2q_F(p) &=& \mathcal{C}_\textrm{Loss}(\chi^q) 
        + \mathcal{C}_\textrm{Loss}(\chi^{\qbar})
        + \frac{2\gamma}{p}\left(\chi^q + \chi^{\qbar} - 2\chi^g \right)
          \left[\frac{1+\np^B}{1-\np^F}\right] \;,\\
0 &=& \mathcal{C}_\textrm{Loss}(\chi^q) 
    - \mathcal{C}_\textrm{Loss}(\chi^{\qbar})
    + \frac{2\gamma}{p}\left(\chi^q - \chi^{\qbar}\right)
                  \left[\frac{1+\np^B}{1-\np^F}\right] \;,
\eeqa
where $\chi^{g,q}=\chiB^{g,q}(p)$ is the off--equilibrium distribution
functions for gluons and quarks, and $q_{I=A,F}$ is the corresponding
source term ($A$ adjoint gluons, $F$ fundamental quarks). The source 
and loss terms are different in the shear ($l=2$) and bulk ($l=0$) 
channels.  In the bulk channel 
\beqa
q_I(p)&\equiv& \left(\frac{1}{3}-c_s^2\right)\frac{p}{T}
   - c_s^2\tilde{m}_I^2\frac{1}{p T}\;,\\
\mathcal{C}_\textrm{Loss}(\chi) &\equiv&  
   \mu_I T\left(-\chi^{\prime\prime} 
     + \left(\frac{1\pm 2\np}{T} - \frac{2}{p}\right)
             \chi^{\prime} \right)\;.
\eeqa
For comparison, we also show the corresponding source and loss term 
in the shear channel,
\beqa
q_I(p) &\equiv& \frac{p}{T}\;,\\
 \mathcal{C}_\textrm{Loss}(\chi) &\equiv& 
  \mu_I T\left(-\chi^{\prime\prime} 
     + \left(\frac{1\pm 2\np}{T} - \frac{2}{p}\right)
        \chi^{\prime} +\frac{6}{p^2}\chi \right)\;.
\eeqa
The coupled second order differential equations for $\chi^{g,q}$ can be 
solved in the same manner as the pure glue case.  The result is shown in 
fig.~(\ref{fig:LL}). We observe that there are important differences
between quarks and gluons, and that there is a shift in the gluon 
distribution due to the presence of quarks. Integrating the distribution 
functions gives a bulk viscosity $\zeta/(T^3\alpha_s^2)\ln = 0.66$ for $N_f=3$.

%%%%%%%%%%%%%%%%%%%%%%%%%%%%%%%%%%%%%%%%%%%%%%%%%%%%%%%%%%%%%%%%%%%%%%%%%
\begin{figure}[t]
\begin{center}
\includegraphics[scale=0.8]{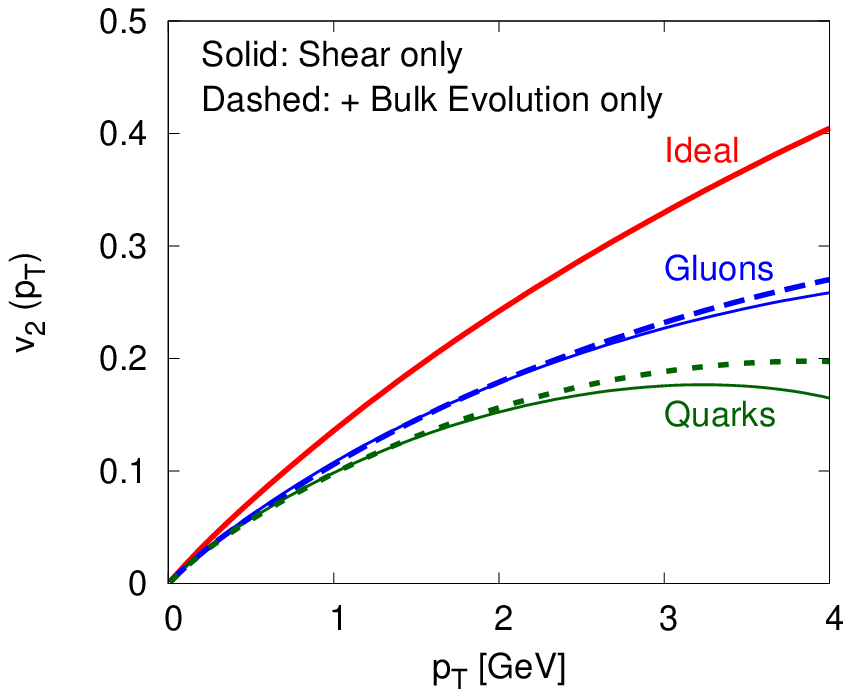}
\includegraphics[scale=0.8]{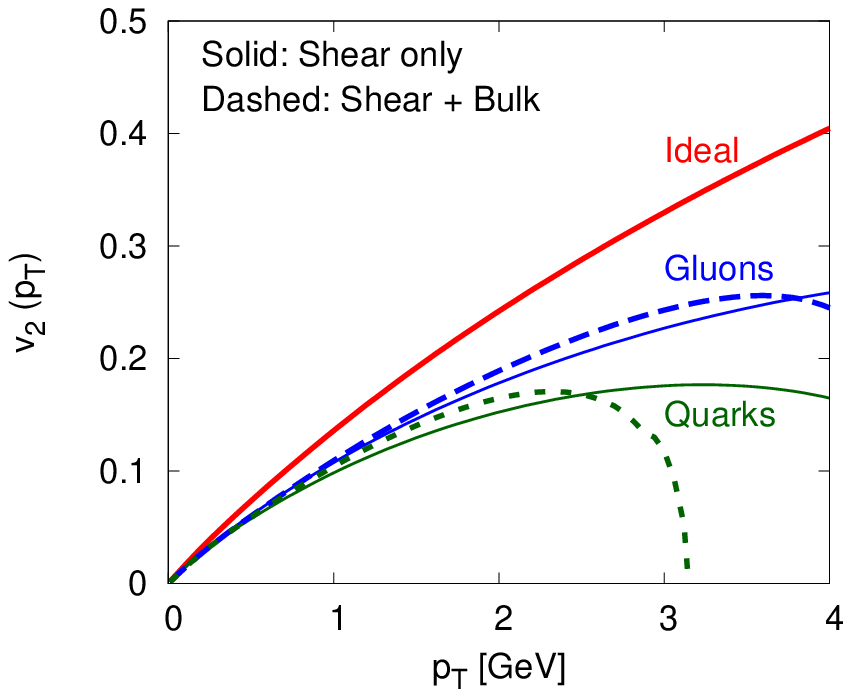}
\caption{Differential elliptic flow of Quarks and Gluons.  The solid 
curves labeled `Quarks' and `Gluons' represent the quark and gluon 
elliptic flow using the leading log form of the shear viscous correction 
to the distribution function.  In both figures the shear viscosity to 
entropy ratio is $\eta/s = 0.16$.  The corresponding dashed curves are 
the results for a viscous hydrodynamic evolution having $\eta/s=0.16$ 
and $\zeta/s = 0.04$.  The dashed curves in the left plot neglect the 
bulk viscous correction to the distribution function at freeze--out.  
The left plot should be taken as strictly pedagogical since energy--momentum 
conservation is violated.  The right plot shows the complete leading log 
result.  Additional details of the hydrodynamic parameters can be found 
in appendix~\ref{app:hydroDetails}.}
\label{fig:v2qgp1}
\end{center}
\end{figure}
%%%%%%%%%%%%%%%%%%%%%%%%%%%%%%%%%%%%%%%%%%%%%%%%%%%%%%%%%%%%%%%%%%%%%%%%%

 We are now in a position to compute viscous corrections to the 
elliptic flow of quarks and gluons. Our calculations are based on 
the 2+1 dimensional second order hydrodynamics code described in 
\cite{Dusling:2007gi}.  See appendix~\ref{app:hydroDetails} for details 
of the hydrodynamic model.  We choose an initial energy density 
appropriate for ${\it Au+Au}$ collisions at 200 AGeV. The results
shown in fig.~\ref{fig:v2qgp1} correspond to an impact parameter 
$b=6.8$ fm. The differential elliptic flow parameter $v_2(p_T)$ 
for quarks and gluons is computed using the strategy outlined 
in the introduction. We have used $m_D=2.9T$ which corresponds to 
$c_s^2=0.2$. For these parameters leading log QCD predicts $\eta/s=0.16$ 
and $\zeta/s=0.08$. These values of the transport coefficients lead
to rather large corrections of the spectra. The results show in 
fig.~\ref{fig:v2qgp1} were obtained for a smaller value of the bulk
viscosity, $\zeta/s=0.04$. 

 In both the left and the right panel of fig.~\ref{fig:v2qgp1} the 
elliptic flow parameter $v_2(p_T)$ in ideal hydrodynamics is shown as
the solid red line, and the elliptic flow of quarks and gluons in a 
simulation with shear viscosity only is shown as the solid green and
blue curves. The dashed curves in the left panel show the result if 
bulk viscosity is included in the hydrodynamic evolution, but not in 
the distribution functions (shear viscosity is included in $\delta f$). 
We note that this procedure violates energy--momentum conservation across 
the freeze--out hyper--surface, but it gives an indication of the role
that bulk viscosity plays in the hydrodynamic evolution.  The inclusion 
of bulk viscosity reduces both the radial flow and the momentum anisotropy.
These two effects lead to a small reduction of $v_2(p_T)$ for $p_T\lsim 2$ 
GeV.

 The right panel in fig.~\ref{fig:v2qgp1} shows the full result including 
the effect of bulk viscosity on the distribution function. Comparing with
the left panel we clearly observe the importance of viscous correction to 
$\delta f$. From eq.~(\ref{eq:del_f}) and fig.~\ref{fig:LL} we can see that 
the shift in the distribution functions due to bulk viscosity is positive 
at small $p_T$.  From fig.~\ref{fig:LL} the sign change in $\chiB$ occurs 
around $p/T \sim 5$. At a decoupling temperature of 150 MeV this corresponds 
to $p_T\lsim 750$ MeV. Taking into account the boost due to radial expansion
the critical $p_T$ is further reduced to $p_T \lsim 400$ MeV, which is 
barely visible on the plot.  At higher momentum the bulk viscosity tends 
to soften the $p_T$ spectra. As the spectra enter into the denominator 
in eq.~(\ref{eq:v2def}) this leads to an increase in $v_2(p_T)$.

 Overall, the effect of bulk viscosity on $v_2(p_T)$ in the regime 
$p_T\lsim 2$ GeV is modest, considering that $\zeta$ is only a factor 
of four smaller than $\eta$.  This result is consistent with the scaling 
relations (\ref{chib_scal}) and (\ref{zeta_scal}). At very weak coupling 
$\zeta$ is suppressed by two powers of the small parameter $(1/3-c_s^2)$, 
whereas $\delta f$ is only suppressed by one power. At strong coupling, 
however, the large numerical coefficient in eq.~(\ref{zeta_scal}) 
enhances $\zeta/\eta$ relative to $\chiB/\chi_\pi$.

%%%%%%%%%%%%%%%%%%%%%%%%%%%%%%%%%%%%%%%%%%%%%%%%%%%%%%%%%%%%%%%%%%%%%%%%%
\subsection{Leading order behavior at large momentum}
%%%%%%%%%%%%%%%%%%%%%%%%%%%%%%%%%%%%%%%%%%%%%%%%%%%%%%%%%%%%%%%%%%%%%%%%%

 In perturbative QCD the leading order result for the bulk viscosity 
is governed by small angle $2\leftrightarrow 2$ scattering, and 
inelastic $2\leftrightarrow 3$ processes are suppressed by a logarithm
of the coupling constant. At large momenta, $p>T/\log(1/g)$, the  
logarithmic suppression is compensated by the growth of the  $2
\leftrightarrow 3$ reaction with energy. In this regime the correction 
to the distribution function is determined by the physics of energy loss. 
Arnold et al.~showed that at leading order in the coupling these 
effects can included in terms of an effective $1\leftrightarrow 2$
collision term \cite{Arnold:2002zm}
\beqa
\frac{ p\nu_a \mathcal{C}_a^{1\leftrightarrow 2}}{(2\pi)^3} =
  \sum_{bc}\int \frac{dx}{x^{5/2}}\mbox{ } 
   \gamma^c_{ab}\big(p;xp,(1-x)p\big) 
   \np^a n^b_{(1-x)\p/x} (1\pm n^c_{\p/x})
      \left[\chi^a_p + \chi^b_{ (1-x)p/x}-\chi^c_{p/x}\right]\nonumber\\
\hspace{1cm}\mbox{}
+\frac{1}{2}\sum_{bc}\int dx \mbox{ }
  \gamma^a_{bc}\big(p;xp,(1-x)p\big) 
   \np^a (1\pm n^b_{x\p}) (1\pm n^c_{(1-x)\p})
   \left[\chi^a_p - \chi^b_{xp}-\chi^c_{(1-x)p}\right]\, ,
\eeqa
where $a,b,c=g,q$ for quarks/gluons and $\chi_p\equiv\chiB(p)$.
The splitting functions $\gamma^a_{bc}$ are given by 
\beqa
\gamma^g_{gg} &=& \sqrt{2 \hat{q}}\,
   \frac{\as C_A d_A}{(2\pi)^4}\sqrt{1+x^2+(1-x)^2}\,
   \frac{1+x^4+(1-x)^4}{(x(1-x))^{3/2}}\\
\gamma^g_{qq}&=&\sqrt{2 \hat{q}}\,
   \frac{\as C_F d_F}{(2\pi)^4}\sqrt{\kappa+x^2+(1-x)^2}\,
   \frac{x^2+(1-x)^2}{(x(1-x))^{1/2}}\\
\gamma^q_{gq} &=& \sqrt{2 \hat{q}}\,
   \frac{\as C_F d_F}{(2\pi)^4}\sqrt{1+\kappa x^2+(1-x)^2}\,
  \frac{1+(1-x)^2}{(x^3(1-x))^{1/2}}
\eeqa
where $\kappa\equiv (2C_F-C_A)/C_A$ and
\beqa
\hat{q}=C_A g^2 T m_D^2 \int\frac{d^2 q_\perp}{(2\pi)^2}
 \frac{1}{q_\perp^2+m_D^2}=C_A \as T m_D^2 
  \ln\left(\frac{\langle k_T^2\rangle}{m_D^2}\right)\, , 
\eeqa
is the transverse diffusion constant that controls energy loss in a 
quark gluon plasma. We can study the effect of the $1\leftrightarrow 2$ 
splitting term on the solution of the Boltzmann equation in the bulk 
channel at large $p_T$. We find that the asymptotic form of $\chiB$ is 
suppressed relative to the asymptotic solution for $\chi_\pi$ by the 
first power of the conformal symmetry breaking parameter, 
\beqa
\chiB^a(p) = \left(\frac{1}{3}-c_s^2\right)\chi^a_\pi(p) 
  \;\;\;\;  \textrm{for } p \gg T\ln^{-1}(1/g)\;.
\eeqa
The asymptotic form of the gluon distribution in the shear channel
is given by
\beqa
\chi^g_\pi(p)\approx \frac{0.7}{\alpha_s T \sqrt{\hat{q}}} p^{3/2}\;,
\eeqa
where we have used $N_f=0$. The corresponding result for the quark 
distribution, as well as the dependence on the number of flavors, is
given in \cite{Dusling:2009df}.

%%%%%%%%%%%%%%%%%%%%%%%%%%%%%%%%%%%%%%%%%%%%%%%%%%%%%%%%%%%%%%%%%%%%%%%%%
\section{Hadronic Gas}
\label{sec_had}
%%%%%%%%%%%%%%%%%%%%%%%%%%%%%%%%%%%%%%%%%%%%%%%%%%%%%%%%%%%%%%%%%%%%%%%%%

 In the previous section we saw that there are significant differences
between the viscous corrections to the differential elliptic flow of 
quarks and gluons. Of course, the spectra of quarks and gluons are 
not directly observable. In this section we study the question whether 
similar differences are expected in the spectra and $v_2(p_T)$ of 
different hadronic species.

%%%%%%%%%%%%%%%%%%%%%%%%%%%%%%%%%%%%%%%%%%%%%%%%%%%%%%%%%%%%%%%%%%%%%%%%%
\subsection{Low temperature pion gas}
\label{sec_pi}
%%%%%%%%%%%%%%%%%%%%%%%%%%%%%%%%%%%%%%%%%%%%%%%%%%%%%%%%%%%%%%%%%%%%%%%%%

 The bulk viscosity of a pion gas was studied by a number of authors
\cite{Chen:2007kx,FernandezFraile:2008vu,Dobado:2011qu,Lu:2011df}. 
Lu and Moore argued that the system is similar to the scalar field theory 
studied in section \ref{sec_phi4}, and that the bulk viscosity is controlled 
by number changing processes\cite{Lu:2011df}. We will therefore follow the 
discussion leading up to eq.~(\ref{chiB_phi4}) and assume that the 
deviation from equilibrium is governed by the near zero--mode,
\beqa
\chi(p)=\chi_0-\chi_1 E_p\;.
\eeqa
The coefficient $\chi_1$ is determined by Landau matching, and the 
coefficient $\chi_0$ is controlled by the inelastic cross--section,
\beqa
\chi_0=\frac{\beta\mathcal{F}}{4\Gamma_{\textrm{inelastic}}}\;,
\eeqa
where $\mathcal{F}$ as written in eq.~(\ref{eq:F}) is a measure of the 
deviation from conformal behavior. In the case of a pion gas we will 
ignore mean--field effects ($\tilde{m}_\pi=m_\pi$), and take the deviation 
from conformality to be driven by the bare mass of the pion.  In this 
case $\mathcal{F}$ takes the form
\beqa
\mathcal{F} = \intPSba{\pi} \left(\frac{p^2}{3}-c_s^2 \Epa{p}{\pi}^2 \right)
   \np(1+\np)\;.
\eeqa
The total inelastic rate is dominated by the lowest order number changing 
process which is kinematically allowed; $\pi\pi\leftrightarrow \pi\pi\pi\pi$.  
The inelastic cross--section also controls the chemical equilibration rate 
of pions.  The rate at which a pion chemical potential will return to 
equilibrium is given by \cite{Pratt:1999ku}
\beqa
\frac{1}{\tau^{\textrm{chem.}}_{\pi}}
   = \frac{\sum_i \left(\delta n^\pi_i\right)^2\; \Gamma_i}{n_\pi}\;,
\eeqa
where the sum is over all reactions which increase the pion number by 
$\delta n^\pi_i$.  We can therefore make the following identification 
between the bulk viscosity and chemical relaxation time,
\beqa
\zeta = \frac{\mathcal{F}^2}{n_\pi}\tau^{\textrm{chem.}}_{\pi}\;.
\eeqa
If we use classical statistics, which is valid for $m_\pi \gg T$, the phase 
space integrals appearing in $\mathcal{F}$ can be evaluated analytically.
Normalizing the bulk viscosity by the entropy 
density we arrive at the following relationship between the bulk viscosity 
of a low temperature pion gas and the chemical equilibration rate,
\beqa
\frac{\zeta}{s} = \frac{m_\pi}{K_2\;K_3}
   \left(\frac{K_2^2-K_1\;K_3}{3K_3+\beta m_\pi K_2 }\right)^2
     \tau^{\textrm{chem.}}_{\pi}\;,
\eeqa
where $K_{i=1,2,3}$ is the modified Bessel function of order $i=1,2,3$ 
evaluated at $(\beta m_\pi)$. The chemical reaction time arising from 
inelastic pion reactions can be computed in chiral perturbation theory.  
For example, the work of \cite{Goity:1993ik} (see also \cite{Song:1996ik}) 
found $\tau^{\textrm{chem.}}_{\pi}=450$ fm/c at $T=140$ MeV and 
$\tau^{\textrm{chem.}}_{\pi}=120$ fm/c at $T=160$ MeV for a pion mass 
$m_\pi=138$ MeV. Based on these calculations we find $\zeta/s\approx 0.14$ 
at $T=140$ MeV and $\zeta/s\approx 0.03$ at $T=160$ MeV.

%%%%%%%%%%%%%%%%%%%%%%%%%%%%%%%%%%%%%%%%%%%%%%%%%%%%%%%%%%%%%%%%%%%%%%%%%
\subsection{Hadronic resonance gas}
\label{sec_res}
%%%%%%%%%%%%%%%%%%%%%%%%%%%%%%%%%%%%%%%%%%%%%%%%%%%%%%%%%%%%%%%%%%%%%%%%%

 The estimate of $\zeta$ for a pure pion gas is likely to be relevant
only in a relatively small temperature regime. In the regime between 
the freeze--out and the critical temperature many resonances are important.
We will assume that the bulk viscosity of a hadronic resonance gas is 
also dominated by number changing processes. If this is the case we 
may approximate the deviation from equilibrium due to bulk viscosity 
for each hadronic species by the near zero--mode 
\beqa
\delta f^a(\p) = -\np^a(1\pm \np^a)\partial_k u^k
            \left(\chi_0^a-\chi_1\Epa{p}{a}\right)\;,
\label{eq:dfHad}
\eeqa
where $\Epa{p}{a}=\sqrt{p^2+m_a^2}$. The coefficient $\chi_1$ (which is 
the same for all species) is determined by the Landau matching condition
\beqa
\delta \epsilon=0=\sum_a \nu_a \intPSa \Epa{p}{a} \delta f^a(\p)\;,
\eeqa
where $a=\pi,K,\dots$ is a sum of all hadronic species in a resonance gas 
having degeneracy $\nu_a$.  Using the generalization of eq.~(\ref{eq:bulkdf}) 
to a system of multiple species we find
\beqa
\zeta = \sum_a \nu_a \chi_0^a\mathcal{F}^a\;,
\label{eq:zetahad}
\eeqa
where
\beqa
\mathcal{F}^a = \intPSba{a} \left(\frac{p^2}{3}-c_s^2 \Epa{p}{a}^2 \right)
   \np^a(1\pm \np^a)\;.
\label{eq:Fa}
\eeqa
As in the case of a dilute pion gas we neglect mean--field effects and 
assume that the deviation from conformality is related to the bare masses of 
the resonances.  The off--equilibrium distribution in a multi--component 
system is determined by one parameter, $\chi_1$, which is common to 
all species, and $N_{\textrm{species}}$ parameters $\chi_0^a$ that are 
different for each species. The parameter $\chi_1$ is determined by the 
Landau matching condition, and one linear combination of the $\chi_0^a$
can be related to the bulk viscosity. Explicit information on inelastic 
hadronic cross--sections is needed to determine the remaining 
($N_{\textrm{species}}-1$) coefficients.

%%%%%%%%%%%%%%%%%%%%%%%%%%%%%%%%%%%%%%%%%%%%%%%%%%%%%%%%%%%%%%%%%%%%%%%%%
\begin{figure}[t]
\begin{center}
\includegraphics[width=8cm]{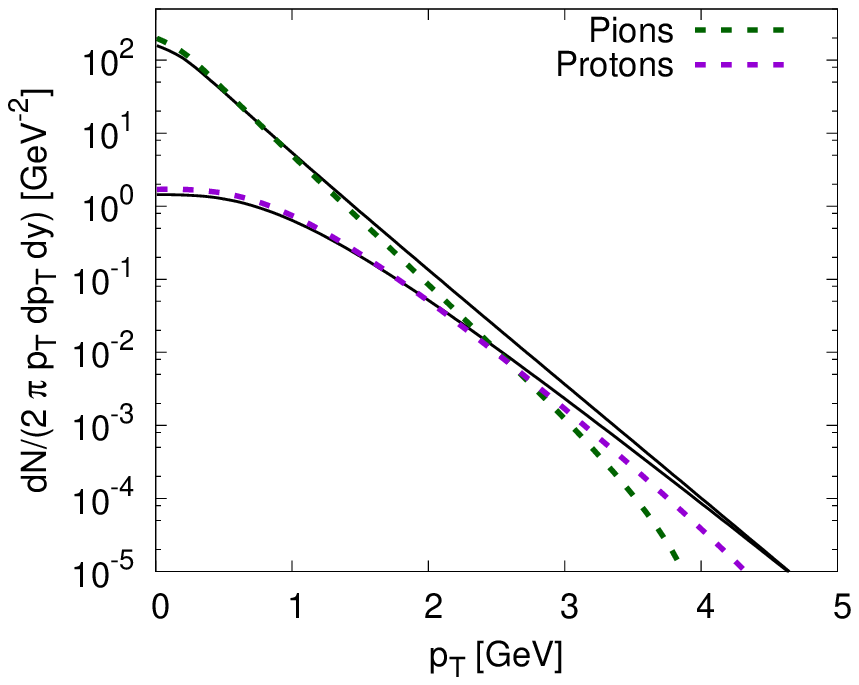}
\includegraphics[width=8cm]{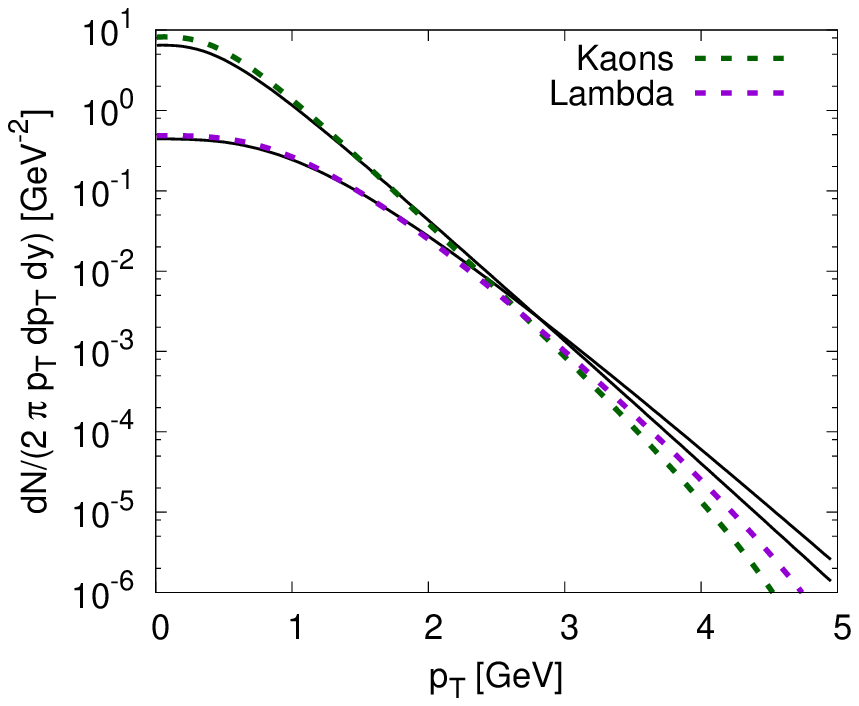}
\caption{Transverse momentum spectra of pions, protons (left panel),
as well as kaons and lambda baryons (right panel). The solid lines
correspond to shear viscosity only, and the dashed lines show the 
result for shear and bulk viscosity with $\eta/s=0.16$ and $\zeta/s
=0.005$. }
\label{fig:pt-spec}
\end{center}
\end{figure}
%%%%%%%%%%%%%%%%%%%%%%%%%%%%%%%%%%%%%%%%%%%%%%%%%%%%%%%%%%%%%%%%%%%%%%%%%

 In this work we will not attempt to compute these inelastic rates. 
Instead, we will rely on a model that is motivated by prior calculations 
of chemical equilibration rates in a hadronic resonance gas 
\cite{Goity:1993ik,Song:1996ik,Pratt:1999ku,Bratkovskaya:2000qy,NoronhaHostler:2007jf}. 
Using a phenomenological model for the inelastic cross--section Pratt 
and Haglin showed that the chemical equilibration time near thermal 
freeze--out is $5-10$ times larger for kaons than it is for pions 
\cite{Pratt:1999ku}. A similar estimate was also obtained in a BUU
transport model \cite{Bratkovskaya:2000qy}.  We therefore expect the 
bulk viscous correction of kaons to be that much larger than pions 
({\em i.e.} $\chi_0^K/\chi_0^\pi \sim 5-10$.). A larger set of resonances 
(but excluding strangeness) was studied by Goity \cite{Goity:1993ik}. 
In this paper the deviation from chemical equilibrium (at fixed 
temperature) is parameterized in terms of effective chemical potentials 
for non-conserved charges like the total number of pions, rho mesons,
nucleons plus anti-nucleons, etc. Goity finds that the largest 
relaxation time corresponds to a chemical potential for meson (baryon) 
resonances approximately twice (2.5 times) larger than that of pions 
near the transition temperature.  

 In the following we will use the ansatz in eq.~(\ref{eq:dfHad}) and
choose $\chi^a_0$ for each meson and baryon species to be a constant 
multiple $C_m$ and $C_b$ of $\chi_0^\pi$, 
\beqa
\chi_0^a=\left\{
\begin{array}{l c}
\chi_0^\pi & \textrm{ Pions}\\
C_m\times \chi_0^\pi & \textrm{ Mesons} \\
C_b\times \chi_0^\pi & \textrm{ Baryons} \\
\end{array}\right.\;.
\eeqa
Due to the strong $\rho \to 2\pi$ reaction rate we expect the $\rho$ 
and $\pi$ mesons to be in relative chemical equilibrium.  This suggests 
that $\mu_\rho = 2\mu_\pi$ and therefore $C_m \approx 2$.  Additionally, 
the average pion multiplicity in the strong $p\overline{p} \to n\pi$ 
reaction is $n\sim 5$ \cite{Rapp:2000gy}, so that $2\mu_N\approx 
5\mu_\pi$ and therefore $C_b \approx 2.5$. These numbers are in good 
agreement with results obtained by Goity \cite{Goity:1993ik}. The 
remaining coefficient $\chi_0^\pi$ is related to the bulk viscosity 
via eq.~(\ref{eq:zetahad})
\beqa
\label{eq:zeta_res}
\zeta=\chi_0^\pi\sum_a\nu_a C_a \mathcal{F}^a\;\;\;\;\textrm{where}\;\;\;\;
C_a=\left\{
\begin{array}{l c}
1 & \textrm{ Pions}\\
C_m & \textrm{ Mesons} \\
C_b & \textrm{ Baryons} \\
\end{array}\right.\;.
\eeqa
We emphasize that in a complete calculation that includes inelastic
rates such as $N\bar{N}\to 5\pi$ the value of $\zeta$ is completely
determined by microscopic dynamics. Without microscopic information
about inelastic rates we can place bounds on $\chi_0^\pi$ from the 
observed spectra, and then extract bounds on $\zeta$ from 
eq.~(\ref{eq:zeta_res}).

%%%%%%%%%%%%%%%%%%%%%%%%%%%%%%%%%%%%%%%%%%%%%%%%%%%%%%%%%%%%%%%%%%%%%%%%%
\begin{figure}[t]
\begin{center}
\includegraphics[scale=0.8]{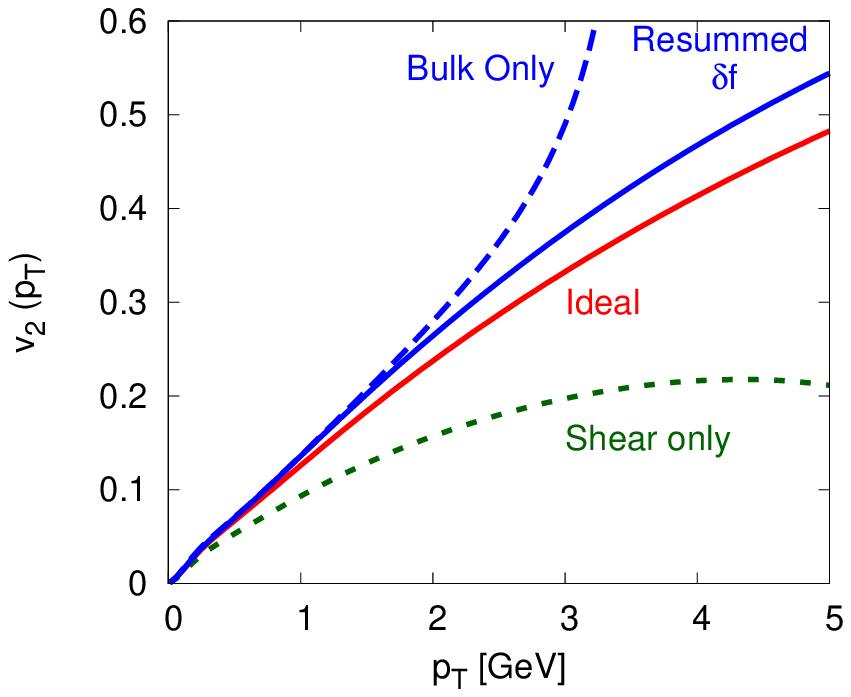}
\includegraphics[scale=0.8]{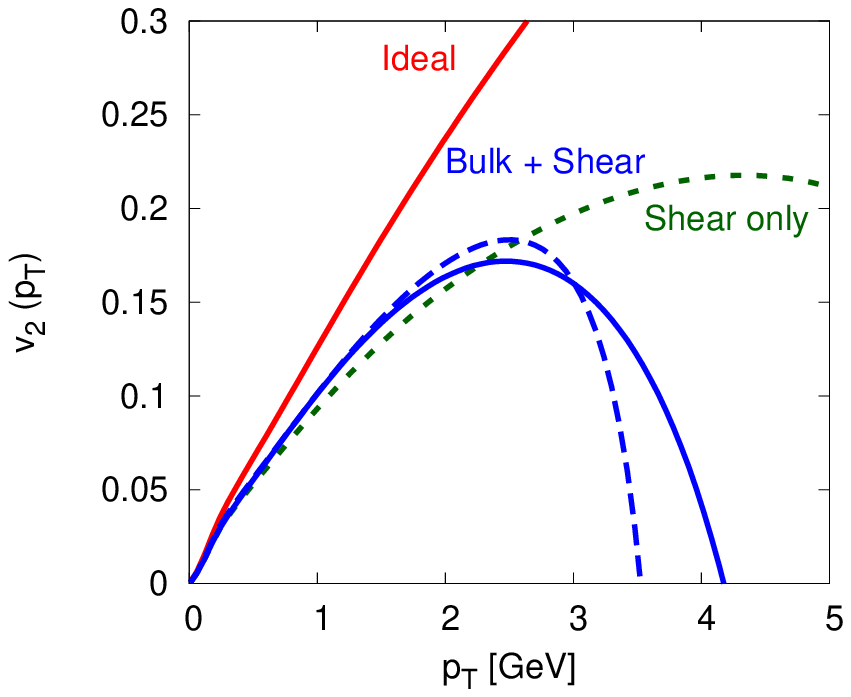}
\caption{The left panel shows the elliptic flow of pions for a bulk viscosity 
at freeze--out of $(\zeta/s)_{\textrm{frzout}}\approx 0.005$.  The dashed 
curve shows the result using the linear form of the viscous correction 
given in eq.~(\ref{eq:dfHad}), and the solid curve shows the result using 
the resummed form given in eq.~(\ref{eq:dfResum}). The right panel shows
the elliptic flow of pions from viscous hydrodynamics when both shear and 
bulk viscosity are included. The two curves labeled `bulk+shear' are
labeled as in the left panel: the dashed line is the linear form of the
distribution function, and the solid line shows the resummed result.}
\label{fig:pions}
\end{center}
\end{figure}
%%%%%%%%%%%%%%%%%%%%%%%%%%%%%%%%%%%%%%%%%%%%%%%%%%%%%%%%%%%%%%%%%%%%%%%%%

 Details of the hydrodynamic simulation are described in appendix
\ref{app:hydroDetails}. We use the same initial conditions and impact 
parameter as in the case of the pure QGP simulation. The equation of 
state  is a parameterization of a lattice QCD equation of state 
\cite{Laine:2006cp}. In the kinetic model defined in eq.~(\ref{eq:dfHad})
we include meson/baryon resonances up to a mass of 1.6 GeV (mesons) and 
1.8 GeV (baryons). We have checked that the corresponding equation of 
state matches the lattice equation of state at freeze--out. Our resonance 
gas model implies $\chi_\pi^0\simeq -100\zeta/(sT)$. We have chosen 
$(\zeta/s)_{\textrm{frzout}}= 0.005$, which corresponds to $\chi_p^0
\simeq -0.5/T$. Using the average expansion rate $(\partial_k u^k)$ at 
freeze--out the value of $\chi_0^\pi$ can be translated into an effective 
pion chemical at freeze--out, see eq.~(\ref{eq:mu_eff}) below. We find 
$\mu_\pi\simeq 25$ MeV. This value is roughly consistent with the pion 
chemical potential $\mu_\pi\simeq 10$ MeV used in the thermal fireball model 
developed by Rapp \cite{Rapp:2000pe}. 

 We note that we use the same speed of sound, and therefore the same 
deviation from conformality, in our calculations in the quark gluon 
plasma phase and the hadron resonance gas. The difference between the 
values of $\zeta/s$ in the two phases is connected with the different 
relations between $\chiB$ and $\zeta$ for the two systems. These relations 
reflect different physical mechanisms for producing bulk viscosity. In 
the the quark gluon plasma bulk viscosity is controlled by momentum 
rearrangement, and shear and bulk viscosity are intimately related, see 
eq.~(\ref{zeta_scal}). In the hadron resonance gas model bulk viscosity 
is dominated by particle number changing processes, and there is no 
direct relation between shear and bulk viscosity. The fairly small value 
of $\zeta/s$ in the hadron resonance gas is further related to cancellations 
between low--mass and high--mass resonances in eq.~(\ref{eq:Fa}). 

%%%%%%%%%%%%%%%%%%%%%%%%%%%%%%%%%%%%%%%%%%%%%%%%%%%%%%%%%%%%%%%%%%%%%%%%%
\begin{figure}[t]
\begin{center}
\includegraphics[scale=.8]{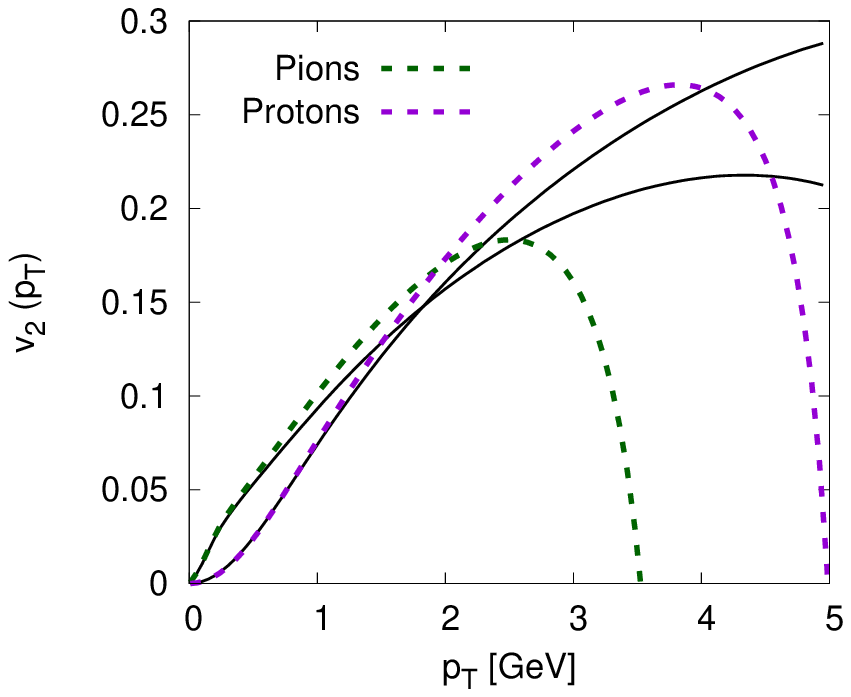}
\includegraphics[scale=.8]{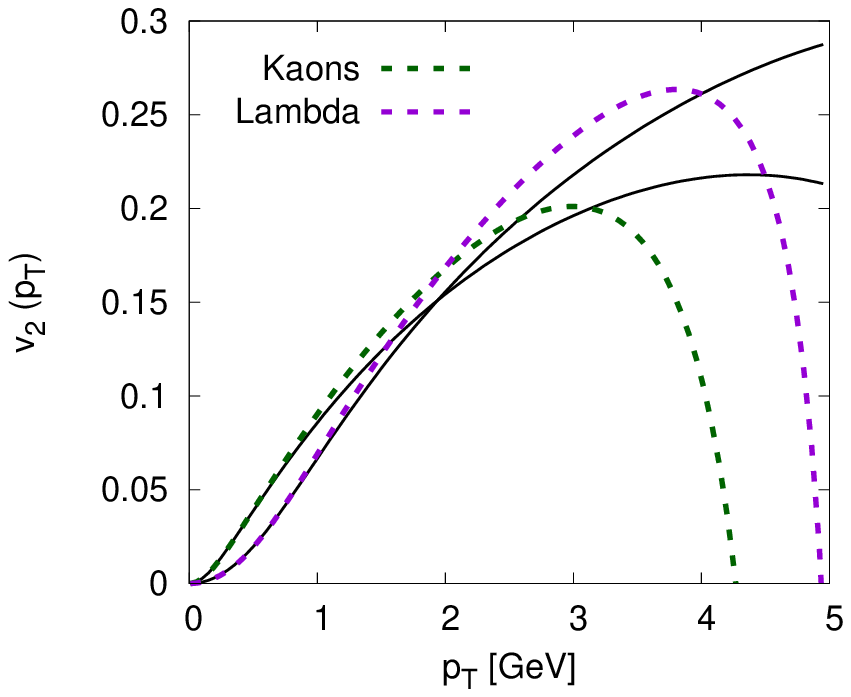}
\caption{Differential elliptic flow $v_2(p_T)$ for pions and protons
(left panel), as well as kaons and lambdas (right panel). The curves
are labeled as in fig.~\ref{fig:pt-spec}. The solid lines show the
result for shear viscosity only, and the dashed lines correspond to
shear and bulk viscosity with $\eta/s=0.16$ and $\zeta/s=0.005$.}
\label{fig:hadrons}
\end{center}
\end{figure}
%%%%%%%%%%%%%%%%%%%%%%%%%%%%%%%%%%%%%%%%%%%%%%%%%%%%%%%%%%%%%%%%%%%%%%%%%

%%%%%%%%%%%%%%%%%%%%%%%%%%%%%%%%%%%%%%%%%%%%%%%%%%%%%%%%%%%%%%%%%%%%%%%%%
\begin{figure}[t]
\begin{center}
\includegraphics[scale=.8]{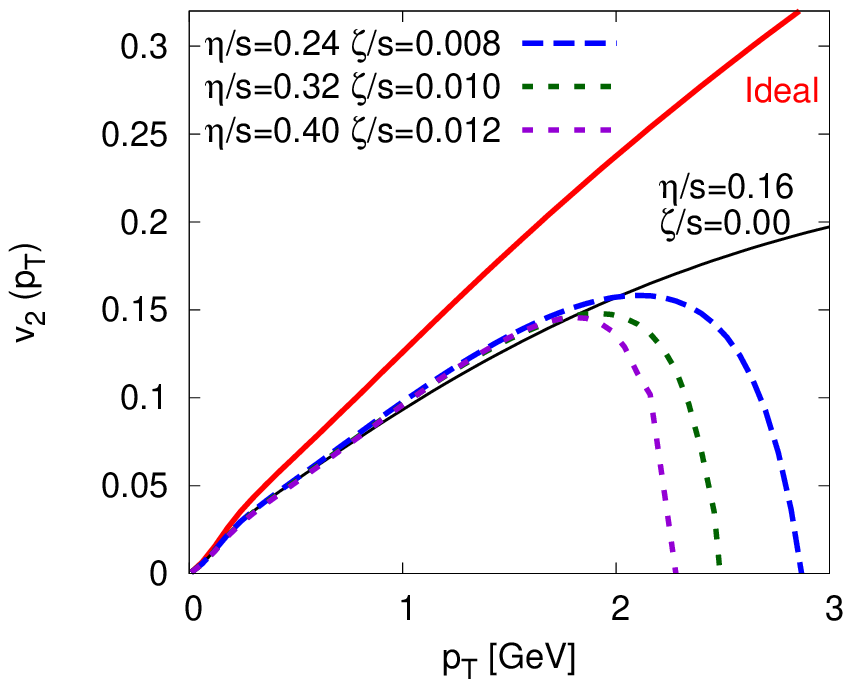}
\includegraphics[scale=.8]{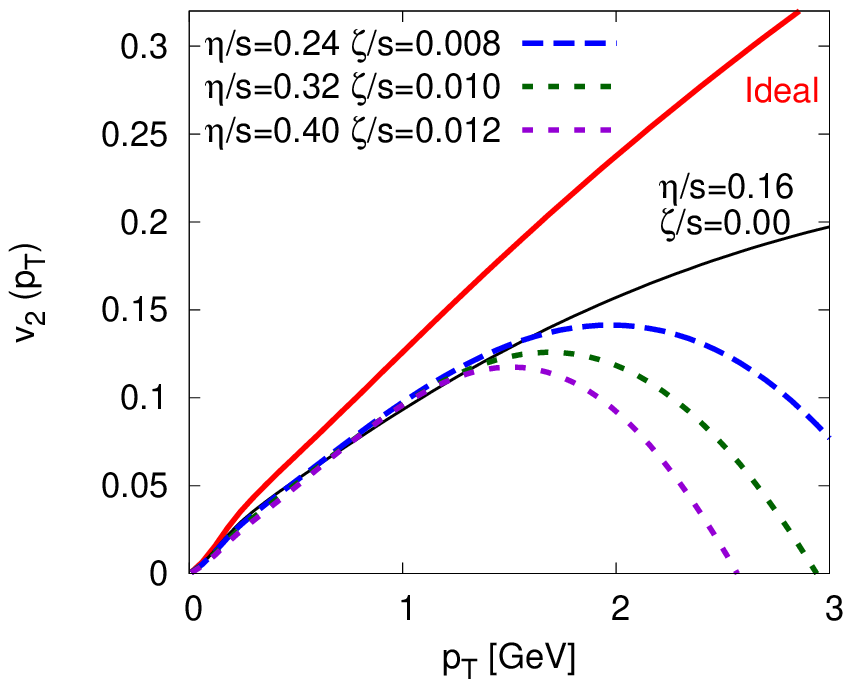}
\caption{Differential elliptic flow of pions using the linearized 
expression for $\delta f$ (left) and the resummed form of $\delta f$ 
(right). The same $v_2(p_T)$ can be obtained for $p_T \lesssim 2$ GeV 
when increasing $\eta/s$ by a factor of 2.5 as long as the bulk 
viscosity is increased as well.}
\label{fig:pionsSum}
\end{center}
\end{figure}
%%%%%%%%%%%%%%%%%%%%%%%%%%%%%%%%%%%%%%%%%%%%%%%%%%%%%%%%%%%%%%%%%%%%%%%%%

In fig.~\ref{fig:pt-spec} we show the $p_T$ spectra of pions, protons,
kaons and lambdas. The shear viscosity was chosen to be $\eta/s=0.16$ as 
in fig.~\ref{fig:v2qgp1}. Corrections to the hadronic spectra due to the 
shear viscosity were computed as described in \cite{Dusling:2009df}.
We observe that, as in the case of quarks and gluons, bulk viscosity 
increases the spectra at small $p_T$, and suppresses the spectra at 
large $p_T$. The high $p_T$ suppression is more prominent in the case 
of pions because the spectra are determined by the competition between 
the constant term $\chi_0^a$ and the linear term $-\chi_1\Epa{p}{a}$ 
term, where the constant contribution is bigger in the case of baryons, 
$\chi_0^B>\chi_0^\pi$. The effect of bulk viscosity on the elliptic 
flow parameter $v_2(p_T)$ is shown in fig.~\ref{fig:pions}. For comparison 
we also show the elliptic flow in the case of an the ideal gas, and in the 
case of shear viscosity only ($\eta/s=0.16$). We find that bulk viscosity 
tends to increase elliptic flow for $p_T\gsim 1$ GeV. The reason is the 
same as in fig.~\ref{fig:v2qgp1}: bulk viscosity suppresses the single 
particle spectra at large $p_T$, and the spectra enter into the denominator 
of the definition of $v_2(p_T)$, see eq.~(\ref{eq:v2def}). The effect 
becomes very large for $p_T\gsim 2.5 $ GeV. A similar behavior was seen 
in \cite{Monnai:2009ad}. Clearly, the large $p_T$ behavior is unphysical 
and stems from the fact that the particle distribution function becomes 
negative at some $p_T$.  In order to circumvent this we can attempt to 
do a resummation of the viscous correction.  We can expand $f^a(\p)$ to 
first order in $\delta T$ and chemical potential $\mu$,
\beqa
\delta f^a(\p) = \np^a(1\pm\np^a)
 \left(\frac{\mu^a}{T}+\frac{\Epa{p}{a}\delta T}{T^2}\right)\, . 
\eeqa
Comparing this with the form of the off--equilibrium distribution given 
in eq.~(\ref{eq:dfHad}) we make the identification
\beqa
\label{eq:mu_eff}
\mu^a&=&-(\partial_k u^k) T\chi_0^a\\
\delta T&=&+(\partial_k u^k) T^2\chi_1
\eeqa
The physics behind this is straightforward.  As a system undergoes an 
expansion (in heavy--ion collisions the expansion rate is $\partial_k u^k 
\sim \frac{1}{\tau}$) the density of the system drops.  However, due to 
the inefficiency of number changing processes there is an excess of 
particles with respect to what would be expected given the energy density 
of the system. This excess of particles can be parameterized by a positive 
shift in the chemical potential. We can resum the viscous correction
by using the ideal distribution function with a shifted temperature 
and chemical potential\footnote{In our calculations we have put 
the factor $e^{\mu^a/T}$ in the numerator in order to avoid possible
problems with Bose condensation in certain regions of phase space.}
\beqa
f^a(\p)\approx \frac{1}
        {e^{\frac{\Epa{p}{a}}{T+\delta T}-\beta\mu^a}\pm 1}\;.
\label{eq:dfResum}
\eeqa
The above non--equilibrium distribution function is manifestly positive 
definite.  The resulting $v_2$ spectrum is shown in the left panel
of fig.~\ref{fig:pions}.  At low $p_T$ the spectrum matches the 
linearized form, but it has the advantage that it is well-behaved 
at high $p_T$.

%%%%%%%%%%%%%%%%%%%%%%%%%%%%%%%%%%%%%%%%%%%%%%%%%%%%%%%%%%%%%%%%%%%%%%%%
\begin{figure}[t]
\begin{center}
\includegraphics[scale=.8]{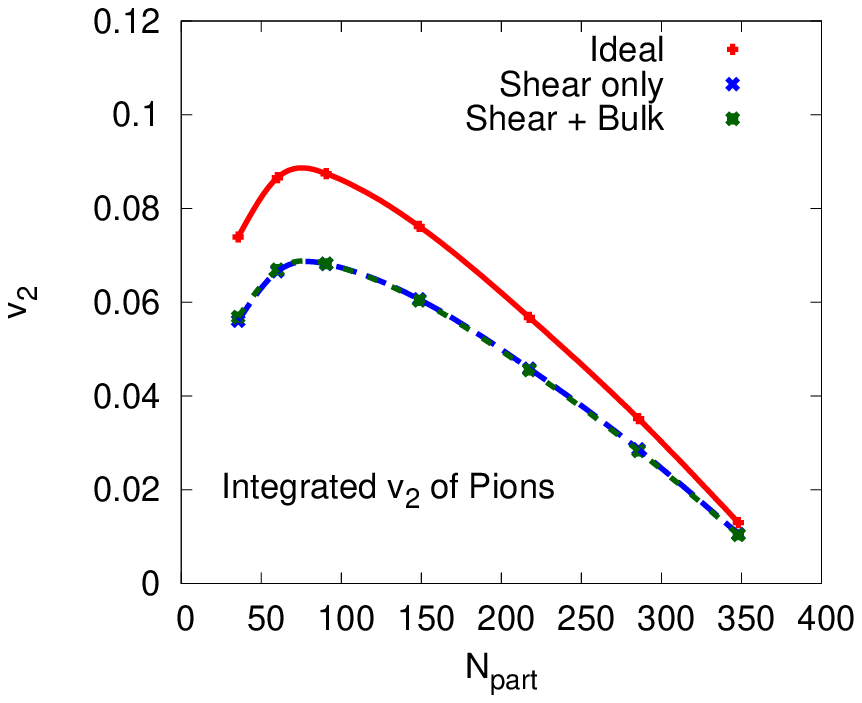}
\includegraphics[scale=.8]{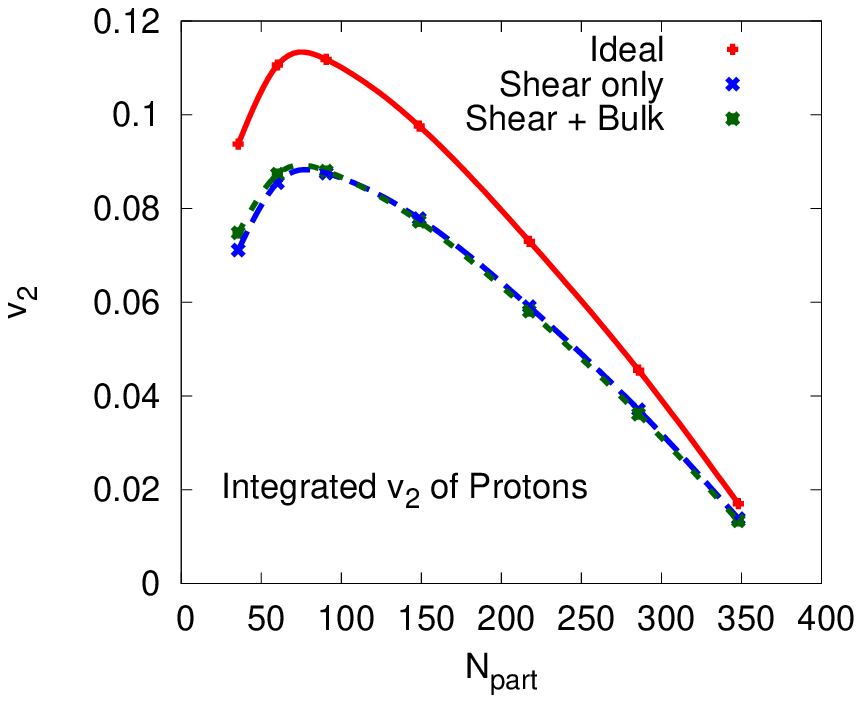}
\caption{Integrated $v_2$ as a function of the number of participants
for pions (left panel) and protons (right panel). We show the result 
in ideal hydrodynamics, the case of only shear viscosity with $\eta/s
=0.16$, and the case of both shear and bulk viscosity with $\eta/s=0.16$
and $\zeta/s=0.005$.}
\label{fig:intv2}
\end{center}
\end{figure}
%%%%%%%%%%%%%%%%%%%%%%%%%%%%%%%%%%%%%%%%%%%%%%%%%%%%%%%%%%%%%%%%%%%%%%%%%

 Resumming the effects of bulk viscosity on the spectra is not as important 
if shear viscosity is also included.  Shear viscosity tends to harden 
the $p_T$ spectra, and therefore prevents the distribution function from 
becoming negative (provided $\eta/s$ is sufficiently large). In the right 
panel of fig.~\ref{fig:pions} we show the elliptic flow of pions when both 
shear and bulk viscosity are taken into account.  In this case we see much 
better agreement between the linear and resummed result even at large $p_T$. 
We observe that the effect of bulk viscosity on the pion $v_2(p_T)$ is 
comparable to the analogous correction to the quark $v_2(p_T)$, despite 
the smaller bulk viscosity used in our simulation of the hadronic phase. 
This is related to the larger numerical coefficient that appears in the 
relation between $\zeta$ and $(\frac{1}{3}-c_s^2)\chi(p)$ in the quark 
gluon plasma compared to the hadron resonance gas. 

 In fig.~\ref{fig:hadrons} we compare viscous corrections to the 
differential elliptic flow parameter $v_2(p_T)$ for different hadronic 
species. Reference \cite{Dusling:2009df} observed that a simple model for 
elastic meson and baryon cross section reproduces the empirically observed 
quark number scaling of $v_2(p_T)$. Fig.~\ref{fig:hadrons} shows that bulk 
viscosity leads to significant modifications of the $v_2(p_T)$ of individual 
species, but the scaling relations between different species are approximately 
preserved.

 At a fixed deviation from conformality the off-equilibrium correction
to the spectrum increases linearly with the bulk viscosity coefficient
$\zeta$. This means that the value of $\zeta$ cannot be increased
by very much without resulting in spectra and flow parameters that 
are in clear disagreement with the data. However, because of the 
partial cancellation between shear and bulk corrections, it is possible
to increase both $\eta$ and $\zeta$ simultaneously without changing
$v_2(p_T)$ very much. This is demonstrated in fig.~\ref{fig:pionsSum},
where we show that $v_2(p_T\lsim 2\,{\rm GeV})$ is fairly stable in
the range $(\eta/s,\zeta/s)=(0.16,0.005)$ to  $(\eta/s,\zeta/s)=
(0.4,0.012)$.

 This result does not imply that the data do not constrain $\eta$ and
$\zeta$ separately. In fig.~\ref{fig:intv2} we show the $p_T$ integrated
flow parameter $v_2$ for pions and protons as a function of the number 
of participants. The number of participants was determined from the 
Glauber model used in \cite{Dusling:2007gi}. We observe that $p_T$
integrated $v_2$ is quite insensitive to the bulk viscosity. There
are two reasons for this result. First, for values of $\zeta/s$ in the 
range studied in this work the effect of bulk viscosity on the velocity 
field is small. Larger values of $\zeta/s$ may lead to stronger effects
on the integrated $v_2$. Second, because of Landau matching, the $p_T$ 
integrated change in the distribution function is small.

%%%%%%%%%%%%%%%%%%%%%%%%%%%%%%%%%%%%%%%%%%%%%%%%%%%%%%%%%%%%%%%%%%%%%%%%%
\section{Summary and Outlook}
%%%%%%%%%%%%%%%%%%%%%%%%%%%%%%%%%%%%%%%%%%%%%%%%%%%%%%%%%%%%%%%%%%%%%%%%%

\begin{figure}[t]
\begin{center}
\includegraphics[scale=1.0]{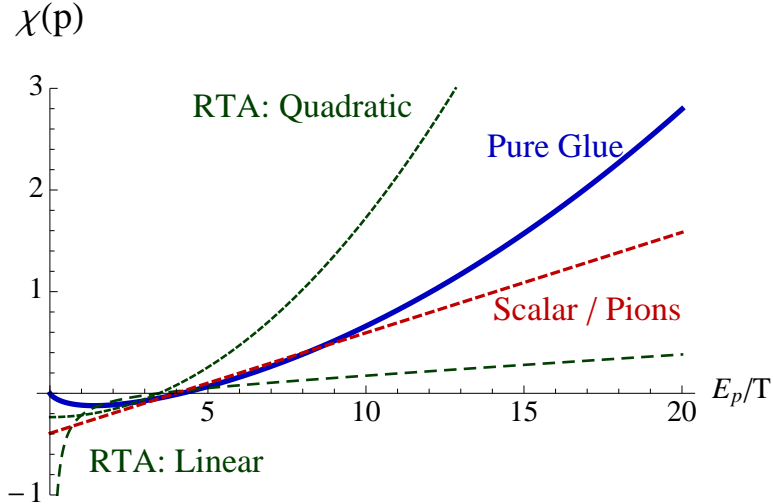}
\caption{In this figure we summarize different functional forms of 
the correction to the single particle distribution function due
to bulk viscosity, $\chiB(p)$. The curves show the linear and 
quadratic form of the relaxation time approximation, the result
in leading log pure gauge theory, and the result in a gas of massive
pions.}
\label{fig:chiSummary}
\end{center}
\end{figure}

 In this work we examined the functional form of the non--equilibrium 
correction to the particle phase--space distribution caused by bulk 
viscosity, see the summary in fig.~\ref{fig:chiSummary}. In the high 
temperature quark-gluon phase the distribution function can be computed 
using the leading log approximation. In this limit bulk viscosity is 
controlled by $2\leftrightarrow 2$ processes that rearrange momentum. 
Particle number changing $2\leftrightarrow 3$ processes only play an
indirect role, in that they prevent the development of an effective 
chemical potential for gluon or quark number. 

 We showed that there is a significant bulk viscous correction to the 
quark and gluon elliptic flow even for a fairly small bulk viscosity 
coefficient.  In addition there are non--trivial differences in the 
quark and gluon off--equilibrium distribution function. These differences
are related to differences in the transport coefficients and effective 
masses. While the quark and gluon distributions are not directly observable,
these distributions serve as direct input for calculations of photon 
and dilepton production from a bulk viscous medium. The effect of shear
viscous corrections to the distribution function on photon and dilepton
production was studied in \cite{Dusling:2008xj,Dusling:2009bc,Bhatt:2010cy,Dion:2011vd}. It is
conceivable that bulk viscosity is responsible for the large elliptic 
flow of photons as compared to hadrons that was recently observed by
the PHENIX collaboration \cite{Adare:2011zr}. This possibility is 
related to the fact that the bulk strain is larger at early times, when 
most photons are produced, and to our observation that bulk viscosity 
enhances $v_2(p_T)$ at intermediate $p_T$.

 For the hadron resonance stage near $T_c$ the calculation of the 
distribution functions is more difficult, and one has to rely on 
simplified models. The simplest model is the relaxation time 
approximation. The relaxation time approximation correctly 
captures the scaling of $\zeta$ and $\chi$ with the deviation
from conformal symmetry, but it cannot predict the functional
form of $\chi(p)$ (it relates the behavior of $\chi(p)$ to the 
unknown energy dependence of $\tau$), and it is in general not
consistent with Landau matching. The relaxation time approximation
also assumes that shear and bulk viscosity are related to the same
process, which need not be the case.

%%%%%%%%%%%%%%%%%%%%%%%%%%%%%%%%%%%%%%%%%%%%%%%%%%%%%%%%%%%%%%%%%%%%%%%%%
%%%%%%%%%%%%%%%%%%%%%%%%%%%%%%%%%%%%%%%%%%%%%%%%%%%%%%%%%%%%%%%%%%%%%%%%%

 A simple model for theories in which bulk viscosity is controlled
by chemical non-equilibration is scalar $\phi^4$ theory. In this 
theory the form of the non-equilibrium distribution functions is 
determined by the exact (energy) and approximate (particle-number) 
zero modes of the collision operator, $\chi\simeq\chi_0-\chi_1 E_{\bf p}$. 
The coefficient of $\chi_0$ is related to the chemical equilibration
time $\tau^{\textrm chem.}$, and $\chi_1$ is fixed by Landau matching.
For a given expansion rate $(\partial^k u_k)=1/\tau$  we can also 
relate $\chi^0$ to the effective chemical potential that describes
the over-population of the single particle distribution function, 
$\mu \simeq  -\frac{T}{\tau}\chi_0$.
 
 The bulk viscosity and non-equilibrium distribution function in 
a low-temperature pion gas is correctly captured by the physics of 
scalar $\phi^4$ theory with the appropriate chemical equilibration time. 
In this work we assume that this is also true for a hadron resonance gas. 
We assume, in particular, that the non-equilibrium distribution 
function of the hadron species $a$ is of the form $\chi^a \simeq 
\chi_0^a-\chi_1 \Epa{p}{a}$, where $\chi_1$ is again fixed by 
Landau matching. The relative magnitude of the coefficient $\chi_0^a$
for different species was fixed by a simple model for the 
effective chemical potentials of meson and baryon resonances. 

 In an expanding system inefficiencies in particle number changing 
processes lead to a particle excess, and both $\chi_0(\partial^k
u_k)$ and $\chi_1 E_{\bf p}(\partial^ku_k)$ are negative. This 
means that bulk viscosity softens the $p_T$ spectra of the produced 
particles. The change in the spectra leads to an enhancement 
of $v_2(p_T)$ at intermediate momenta $p_T\sim (1-2)$ GeV. 

 This enhancement tends to cancel against the effects of shear viscosity. 
We showed, however, that the shear viscosity can be determined reliably 
by focusing on the $p_T$ integrated elliptic flow parameter. We also 
showed that bulk viscosity tends to preserve the approximate ``quark 
number scaling'' observed in in the identified particle $v_2(p_T)$. Once 
$\eta$ is fixed bulk viscosity is strongly constrained by the spectra and 
$v_2(p_T)$. The main difficulty is that in the hadron resonance gas the 
relationship between $\chi(p)$ and $\zeta$ is very sensitive to the 
contribution from high lying resonances. 

%%%%%%%%%%%%%%%%%%%%%%%%%%%%%%%%%%%%%%%%%%%%%%%%%%%%%%%%%%%%%%%%%%%%%%%%%
\begin{figure}[t]
\begin{center}
\includegraphics[scale=1.0]{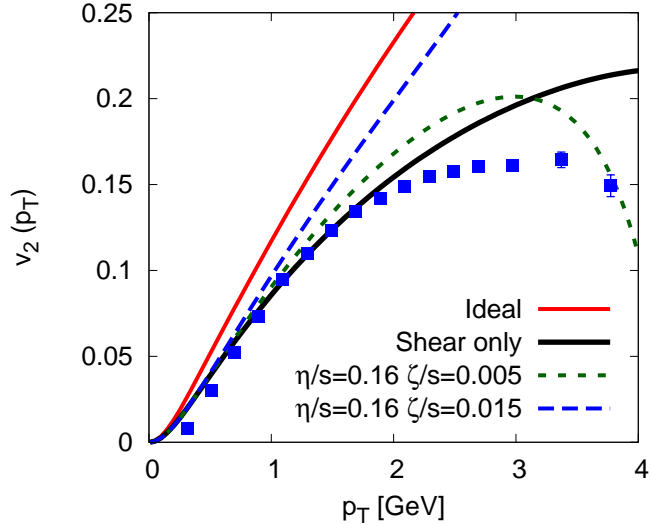}
\caption{Elliptic flow of $K_S$ mesons from viscous hydrodynamics.  The 
hydrodynamic model was tuned such that the ``shear only" result (solid 
black curve) fits the data points. The short--dashed green curve and 
long--dashed blue curve show results from viscous hydrodynamics having a 
bulk viscosity to entropy ratio $\zeta/s=0.005$ and $\zeta/s=0.015$,
respectively. The data were obtained by the STAR collaboration at RHIC
\cite{:2008ed}.}
\label{fig:Ks}
\end{center}
\end{figure}
%%%%%%%%%%%%%%%%%%%%%%%%%%%%%%%%%%%%%%%%%%%%%%%%%%%%%%%%%%%%%%%%%%%%%%%%%

 For the results shown In figs.~\ref{fig:pt-spec}-\ref{fig:intv2} we 
used $(\zeta/s)_{\textrm{frzout}} \lsim 0.005$, and found modest bulk viscous 
correction to $v_2(p_T)$. In order to obtain a rough bound on the maximum
value of $\zeta/s$ allowed by the data obtained at RHIC we have studied
the dependence of our results on $\zeta/s$. Figure~\ref{fig:Ks} shows 
the $v_2(p_T)$ for identified $K_S$ mesons. We have chosen $K_s$ mesons 
because the contribution from resonance decays, which were not included 
in this work, are negligible.  Our hydrodynamic model was tuned previously 
to reproduce the measured spectra using shear viscosity only. This implies 
that the inclusion of bulk viscosity will typically worsen the agreement 
with data.  For $(\zeta/s)_{\textrm{frzout}} = 0.005$ discrepancies with the
data are not large, and the previous level of agreement could presumably
be restored by retuning the parameters of the hydrodynamic model.  For
$(\zeta/s)_{\textrm{frzout}} \approx 0.015$ the discrepancy with data in the 
range $1 \lsim p_T \lsim 2$ GeV is significant, and it is unlikely that 
agreement with the data could be achieved without affecting other observables, 
like the $p_T$ integrated $v_2$.  We therefore feel that it is safe to 
claim that the resonance gas model implies $(\zeta/s)_{\textrm{frzout}} \lsim 
0.015$. We plan to perform more detailed fits in the future. 

 The most important uncertainty in this bound is related to model 
dependence in the relation between $\chi(p)$ and $\zeta$. In the hadron 
resonance gas this relation depends on the inelastic cross--sections of 
high lying resonances. We can estimate the uncertainty of our results by 
reducing the number of resonances included in the model. For example, if 
we only keep mesons (baryons) with masses below 0.8 (1.0) GeV we find 
$\chi_\pi^0\simeq -30\zeta/(sT)$. This relation allows for roughly identical 
fits to the  spectra with a $\zeta/s$ larger by about a factor of three. We 
conclude that a more conservative bound is given by $(\zeta/s)_{\textrm{frzout}} 
\lsim 0.05$.  We emphasize that the data support a non--vanishing bulk 
viscosity. Statistical fits to hadronic yields \cite{Letessier:2005qe} 
show the need to increase the abundance of baryons ({\em i.e.} protons 
+ anti--protons) through a chemical--abundance factor\footnote{The abundance 
factor $\gamma$ has to be distinguished from the fugacity $\lambda=e^{\mu/T}$ 
which enhances the abundance of particles while suppressing that of 
anti--particles.} $\gamma_q \approx 1.6$ at RHIC energies.  This 
result can be naturally accounted for in terms of a non--vanishing 
bulk viscosity.  

 There are a number of issues that we have not addressed in this work. 
Clearly, more work is needed to constrain the bulk viscosity of a 
hadron resonance gas. We have also not taken into account a possible 
increase in the bulk viscosity near $T_c$ due to critical fluctuations 
\cite{Kharzeev:2007wb,Moore:2008ws}. If there is a rapid increase in 
the bulk viscosity near $T_c$ one also expects a rapid rise in the bulk 
relaxation time.  Onuki \cite{PhysRevE.55.403} showed that the bulk 
relaxation time diverges near $T_c$ more rapidly than the bulk viscosity.  
This implies that the system may free--stream through the transition 
region without significant effects on single particle observables.
Clearly, further study in this direction is necessary.

Acknowledgments: KD would like to thank Daniel Fernandez-Fraile for useful 
discussions.  This work was supported by the US Department of Energy grant 
DE-FG02-03ER41260. 

\appendix
%%%%%%%%%%%%%%%%%%%%%%%%%%%%%%%%%%%%%%%%%%%%%%%%%%%%%%%%%%%%%%%%%%%%%%%%%
\section{Details of the hydrodynamic evolution}
\label{app:hydroDetails}
%%%%%%%%%%%%%%%%%%%%%%%%%%%%%%%%%%%%%%%%%%%%%%%%%%%%%%%%%%%%%%%%%%%%%%%%%

%%%%%%%%%%%%%%%%%%%%%%%%%%%%%%%%%%%%%%%%%%%%%%%%%%%%%%%%%%%%%%%%%%%%%%%%%
\begin{figure}[t]
\begin{center}
\includegraphics[scale=.8]{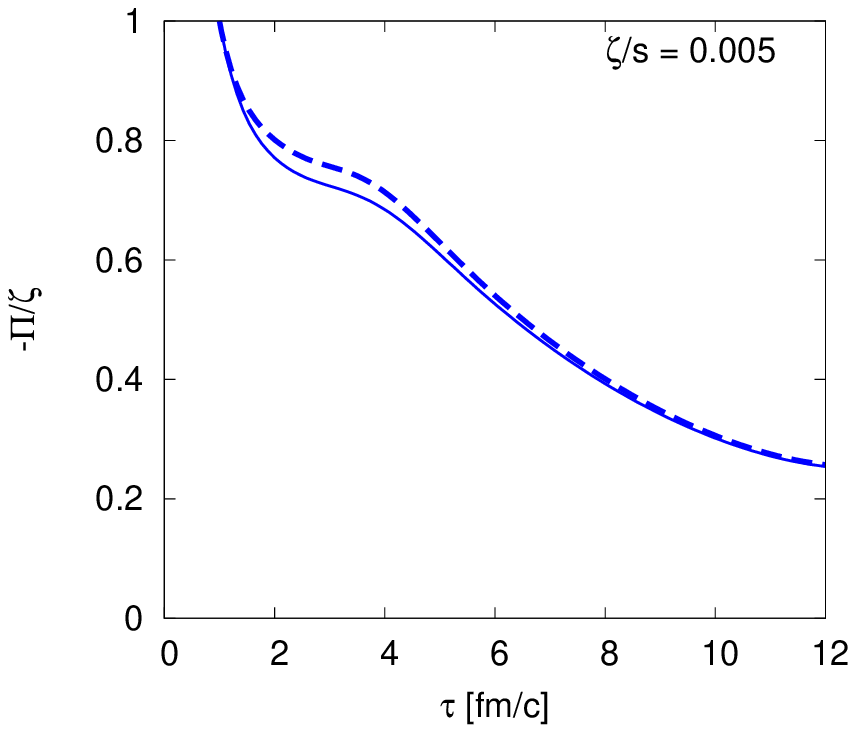}
\includegraphics[scale=.8]{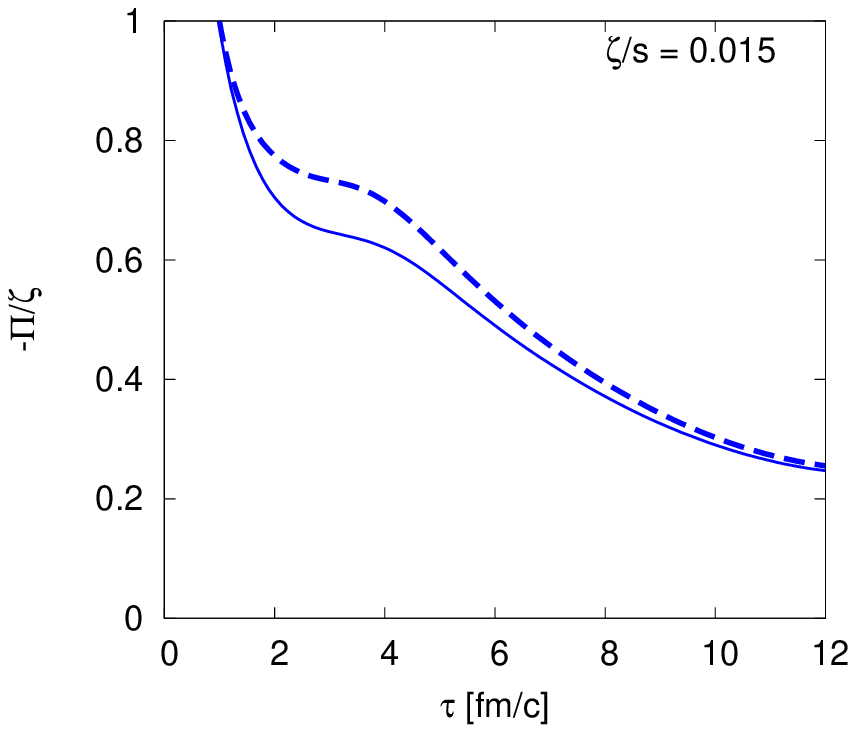}
\caption{Bulk viscous pressure $-\Pi/\zeta$ (solid curves) versus 
proper time along the freeze--out hypersurface shown against the 
Navier--Stokes value $\partial_k u^k$ (dashed curve) for 
$(\zeta/s)_{\textrm{frzout}}\approx 0.005$ (left) and 
$(\zeta/s)_{\textrm{frzout}}\approx 0.015$ (right).}
\label{fig:grad}
\end{center}
\end{figure}
%%%%%%%%%%%%%%%%%%%%%%%%%%%%%%%%%%%%%%%%%%%%%%%%%%%%%%%%%%%%%%%%%%%%%%%%%

  In this appendix we summarize some details of the hydrodynamic 
calculations that were used to compute the velocity and temperature 
profiles that determine the spectra of produced particles. We 
assume longitudinal boost invariance with initial conditions in 
the transverse plane taken from a Glauber Model (see appendix A in 
\cite{Dusling:2009df} for more details). For all non--central 
collisions we have used an impact parameter of $b=6.8$ fm, and 
a decoupling temperature $T_{\textrm{frzout}}=150$ MeV. 

%%%%%%%%%%%%%%%%%%%%%%%%%%%%%%%%%%%%%%%%%%%%%%%%%%%%%%%%%%%%%%%%%%%%%%%%%
\begin{figure}[t]
\begin{center}
\includegraphics[scale=1.0]{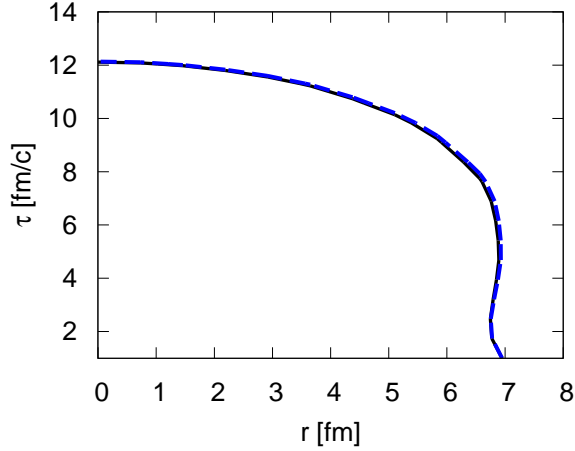}
\caption{Freeze-out hypersurface ($T_{\textrm{frzout}}=150$ MeV) 
for a central ($b=0$) collision with $\sigma_0=0.01$ 
($(\zeta/s)_{\textrm{frzout}}\approx 0.005$) shown as the solid 
black curve and for $\sigma_0=0.03$ ($(\zeta/s)_{\textrm{frzout}}
\approx 0.015$) shown as the dashed blue curve. }
\label{fig:fosurf}
\end{center}
\end{figure}
%%%%%%%%%%%%%%%%%%%%%%%%%%%%%%%%%%%%%%%%%%%%%%%%%%%%%%%%%%%%%%%%%%%%%%%%%

 We solve second order hydrodynamic equations using a second order fluid 
model developed by Grmela and \"Ottinger~\cite{Grmela:1997zz,Ottinger:1998zz}. 
This model is quite similar to the theory of Israel and Stewart
\cite{Israel:1976tn,Israel:1979wp}. Grmela and \"Ottinger introduce 
a new dynamical tensor variable $c_{\mu\nu}$. We will see below 
that this variable is closely related to the velocity gradient 
tensor $\pi_{\mu\nu}$. In the local rest frame the stress energy tensor 
takes the form
\beqa
T^{ij}_{LRF}=p(\delta^{ij}-\alpha c^{ij})\, ,
\eeqa
where $\alpha$ is a small parameter, which will be shown to be 
related to the relaxation time. The tensor variable $c_{\mu\nu}$ is 
conveniently defined to have the property
\beqa
c_{\mu\nu}u^\nu=u_\mu\, .
\eeqa
We decompose $c_{\mu\nu}$ in terms of isotropic and traceless 
components $\overline{c}$ and $\mathring{c}$,
\beqa
c_{\mu\nu}=-u_\mu u_\nu + \mathring{c}_{\mu\nu}+\overline{c}_{\mu\nu}\, , \\
\overline{c}_{\mu\nu}=\frac{1}{3}\left(c^{\lambda}_\lambda-1\right)
 \left(g_{\mu\nu}+u_\mu u_\nu\right)\, .
\eeqa
The equations of motion are dictated by conservation of energy and 
momentum $\partial_\mu T^{\mu\nu}=0$ along with an evolution equation 
for the tensor variable $c_{\mu\nu}$,
\beqa
u^\lambda\left(\partial_\lambda c_{\mu\nu}-\partial_\mu c_{\lambda\nu}
  -\partial_\nu c_{\mu\lambda}\right)
  =-\frac{1}{\tau_0}\overline{c}_{\mu\nu}
   -\frac{1}{\tau_2}\mathring{c}_{\mu\nu}\, ,
\label{eq:cevol}
\eeqa
In the limit that the relaxation times ($\tau_0, \tau_2$) are very 
small the evolution equation yields
\beqa
c^{ij} =\tau_2 \left(\partial_i u^j+\partial_j u^i
            -\frac{2}{3}\delta^{ij}\partial_k u^k\right)
     +\frac{2}{3}\tau_0\delta^{ij}\partial_k u^k\, .
\label{cijequ}
\eeqa
Substituting the above equation into $T^{ij}_{LRF}$ and comparing 
the result to the Navier-Stokes equation the bulk and shear viscosities 
can be identified as
\beqa
\eta=\tau_2 p \alpha \nonumber\, , \\
\zeta=\frac{2}{3}\tau_0 p \alpha\, .
\eeqa
In our work we have taken the parameter $\alpha=0.7$. These relaxation 
times are small enough so that the Navier--Stokes limit is approximately 
maintained near freeze--out.  This is demonstrated in fig.~\ref{fig:grad} 
where the bulk viscous stress $\Pi$ is plotted versus the Navier--Stokes 
expectation for a central ($b=0$) collision.  For reference we also 
show the corresponding freeze--out hypersurface in fig.~\ref{fig:fosurf}.
The dynamical variable $c_{\mu\nu}$ was initialized to the Navier--Stokes 
value.

 In fig.~\ref{fig:v2qgp1} we 
show the elliptic flow of quark and gluons obtained in a simulation 
with a pure QGP equation of state. In order to allow for a speed of 
sound that is different from the conformal value $c_s^2=1/3$ we use a 
polytropic equation of state
\beqa
\mathcal{P}=\left(\gamma-1\right)\epsilon \, . 
\eeqa
The adiabatic index $\gamma$ is chosen in order to fix a constant 
sound speed $c_s^2=0.2$ compatible with lattice parameterizations
near $T_c$. The viscous correction 
to the distribution was computed with a Debye mass $m_D=3.9 T$ so 
that the QGP sound speed is $c_s^2=0.2$, consistent with the speed
of sound used in the hydrodynamic evolution. We employed a simple
parametrization of the solution of the Fokker--Planck equation for 
the off-equilibrium distribution functions. The parametrization
is given in table~\ref{tab:chi}.

%%%%%%%%%%%%%%%%%%%%%%%%%%%%%%%%%%%%%%%%%%%%%%%%%%%%%%%%%%%%%%%%%%%%%%%%%
\begin{table}[t]
\begin{center}\begin{tabular}{|c|cc|}\hline
$$ &  Quarks    &   Gluons   \\
   \hline
$p_0$   & 2.51 & 4.32\\
$c_0$   & $9.56\times 10^{-2}$ & $6.28\times 10^{-2}$ \\
$x_0$   & $5.25\times 10^{-1}$ & $9.56\times 10^{-1}$ \\
$c_1$   & $8.64\times 10^{-3}$ & $3.43\times 10^{-6}$ \\
$x_1$   & 1.66 & 3.48 \\ \hline
\end{tabular}\end{center}
\caption{Parameterization of the leading log QCD off-equilibrium 
distribution function. We use the functional form $\chiB(p)=\left(
c_0 p^{x_0}+c_1 p^{x_1}\right)\ln\left(p/p_0\right)$, for $m_D=3.9$ 
and $N_f=2$.  The above parameterization yields $\zeta/T^3\approx 3.07$.}
\label{tab:chi}
\end{table}
%%%%%%%%%%%%%%%%%%%%%%%%%%%%%%%%%%%%%%%%%%%%%%%%%%%%%%%%%%%%%%%%%%%%%%%%%

 All final state hadron spectra shown in this work were calculated
using a realistic equation of state which is a parameterization of 
the lattice QCD equation of state from \cite{Laine:2006cp}.  This 
equation of state matches on to our hadron resonance gas equation 
of state below $T\sim 160$ MeV.  The bulk viscosity during the 
hydrodynamic evolution was assumed to scale with the second power
of conformality breaking, 
\beqa
\zeta/s=15 \sigma_0 \left(\frac{1}{3}-c_s^2\right)^2\, , 
\label{eq:bulksig0}
\eeqa
where $\sigma_0$ is a free parameter chosen to set the desired magnitude 
of the bulk viscosity coefficient near freeze--out.  At our freeze--out 
temperature of 150 MeV the lattice equation of state used in this work 
yields $c_s^2\approx 0.15$.   In section \ref{sec_res} we examine a 
hadronic resonance gas with  $(\zeta/s)_{\textrm{frzout}}\approx 0.005$, 
corresponding to $\sigma_0=0.01$.

%%%%%%%%%%%%%%%%%%%%%%%%%%%%%%%%%%%%%%%%%%%%%%%%%%%%%%%%%%%%%%%%%%%%%%%%%
\section{Phase Space Integrals}
%%%%%%%%%%%%%%%%%%%%%%%%%%%%%%%%%%%%%%%%%%%%%%%%%%%%%%%%%%%%%%%%%%%%%%%%%

%%%%%%%%%%%%%%%%%%%%%%%%%%%%%%%%%%%%%%%%%%%%%%%%%%%%%%%%%%%%%%%%%%%%%%%%%
\subsection{Relaxation time approximation}
\label{app:RTA}
%%%%%%%%%%%%%%%%%%%%%%%%%%%%%%%%%%%%%%%%%%%%%%%%%%%%%%%%%%%%%%%%%%%%%%%%%

In the relaxation time approximation we found the relationship between 
the shear viscosity and energy--dependent relaxation time $\tau_R(\Ep)$ 
in eq.~(\ref{eq:etaRTA}) which we rewrite here
\beqa
\eta = \frac{\beta}{30\pi^2}\int \frac{p^6}{\Ep^2}\tau_R(\Ep) 
  \np(1\pm \np)\,dp\;.
\eeqa
If we take a relaxation time of the form
\beqa
\tau_R(\Ep)=\tau_0\beta\left(\beta \Ep\right)^{1-\alpha}\;,
\eeqa
the relationship becomes
\beqa
\eta = \tau_0 \frac{\beta^4}{30\pi^2}\int \frac{p^6}{(\beta\Ep)^{1+\alpha}} 
   \np(1\pm \np)\,dp\;.
\label{eq:RTA39}
\eeqa
Making the change of variables $x\equiv \beta\Ep$ we find 
eq.~(\ref{eq:etaRTA2})
\beqa
\eta = \tau_0 \frac{T^3}{30\pi^2}\mathcal{I}_\alpha(\beta m)
\eeqa
where the remaining phase space integral is
\beqa
\mathcal{I}_\alpha(\beta m)\equiv \int_{\beta m}^\infty 
 \frac{\left(x^2-(\beta m)^2\right)^{5/2}}{x^\alpha} \nx(1\pm \nx) dx\, . 
\eeqa
Even though we have arrived at the above phase space integral by studying 
the relaxation time approximation, it will turn out we will need the same 
phase space integrals in other contexts as well.  It is therefore worthwhile 
to study some limits where analytic results can be obtained.  For or a 
classical gas we can replace $\nx(1\pm \nx)\to \nx$ and the phase space
integral can be computed analytically when $\alpha=0$ 
\beqa
\mathcal{I}_{\alpha=0}=15 (\beta m)^3 K_3(\beta m)
\eeqa
Another case where an analytic expression can be found is in the high 
temperature limit ($\beta m\to 0$). For $\alpha < 4$ we find  
\beqa 
\mathcal{I}_\alpha(\beta m = 0)=\G(6-\alpha)
\eeqa 
where for convenience we have defined\footnote{We have used the relation
\beqa
\int_0^\infty\frac{x^{n-1}}{e^x\mp 1}dx=\zeta_{\pm}(n)\Gamma(n)
\eeqa
which can be derived by expanding the numerator in terms of its geometric 
series and then performing the integral of each term in the series 
individually. The remaining summation will then be of the form \ref{eq:zeta}.}
\beqa
\G(x)\equiv \left\{
\begin{array}{l c}
\Gamma(x) & \textrm{Maxwell}\\
\Gamma(x)\zeta_+(x-1) & \textrm{Bose} \\
\Gamma(x)\zeta_-(x-1) & \textrm{Fermi} \\
\end{array}\right. \, , 
\eeqa
for $x>2$ and where 
\beqa
\zeta_{\pm}(s)\equiv \sum_{k=1}^{\infty}\frac{\left(\pm\right)^{k-1}}{k^s}\;.
\label{eq:zeta}
\eeqa
$\zeta_+(s)$ is the usual Riemann--Zeta function and $\zeta_-(s)=(1-2^{1-s})
\zeta_+(s)$. We will also need the $\alpha=4$ behavior of of the above phase 
space integral.  For classical and Fermi statistics the above results hold 
as long as we note that $\lim_{s\to 1} \zeta_-(x)=\ln 2$.  We therefore have 
that $\mathcal{I}_{\alpha=4}=\Gamma(2)$ for classical statistics and 
$\mathcal{I}_{\alpha=4}=\Gamma(2)\ln(2)$ for Fermi statistics.  The above 
integral is logarithmically divergent for bosons when $\alpha=4$.  The 
divergence is regulated by the mass (or thermal mass) of the relevant 
quasi--particles.  We define the following values for $\G(x=2)$ 
\beqa
\G(x=2)\equiv\left\{
\begin{array}{l c}
\Gamma(2) & \textrm{Maxwell}\\
\ln\left(\frac{2T}{m}\right)-\frac{8}{15} & \textrm{Bose} \\
\Gamma(2)\ln(2) & \textrm{Fermi} \\
\end{array}\right. \, . 
\eeqa
The relevant phase space integral for bulk viscosity can be found by using 
the change of variable $x\equiv \beta\Ep$ in eqs.~(\ref{eq:BP1}) and 
(\ref{eq:BP2}),
\beqa
\mathcal{J}_\alpha(\beta m,\beta\tilde{m}) \equiv 
 \int_{\beta m}^\infty \frac{\left(x^2-(\beta m)^2\right)^{5/2}}{x^\alpha} 
  \nx(1\pm \nx)\left[\frac{1}{3}-c_s^2
   \left(1+\frac{(\beta \tilde{m})^2}{x^2-(\beta m)^2}\right)\right]^2 dx\;.
\eeqa
As in the shear case analytic expressions are available. 
For $\alpha=0$ and classical statistics we find
\beqa
\mathcal{J}_{\alpha=0} &=&
  15 \left(\frac{1}{3}-c_s^2\right)^2\;(\beta m)^3 K_3(\beta m) 
 -6 (\beta \tilde{m}c_s)^2\left(\frac{1}{3}-c_s^2\right)
         \;(\beta m)^2 K_2(\beta m)\nonumber\\[0.1cm]
   & & \mbox{} +  (\beta \tilde{m}c_s)^4\;(\beta m)K_1(\beta m)\, . 
\eeqa
If both $\beta m$ and $\beta \tilde{m}$ are taken to zero the integral is
\beqa
\mathcal{J}_{\alpha}&=&\left(\frac{1}{3}-c_s^2\right)^2\;\G(6-\alpha)
\eeqa
Another limit of interest is when $(\beta m)\to 0$ but $\tilde{m}$ 
remains finite.  This is physically relevant since $\tilde{m}$ quantifies 
the deviations from conformality, which is crucial to keep when studying 
bulk viscosity, while the bare or thermal mass only effects the kinematics 
in the phase space integrals.  The only subtlety is if the phase space 
integral is logarithmically divergent in which case the mass serves as 
a cutoff for the integral.  The resulting expression in this limit is
\beqa
\mathcal{J}_{\alpha} = 
  \left(\frac{1}{3}-c_s^2\right)^2\;\G(6-\alpha)
 - 2(\beta \tilde{m}c_s)^2\left(\frac{1}{3}-c_s^2\right)\;\G(4-\alpha)
 +  (\beta \tilde{m}c_s)^4\;\G(2-\alpha)\;.
\eeqa

%%%%%%%%%%%%%%%%%%%%%%%%%%%%%%%%%%%%%%%%%%%%%%%%%%%%%%%%%%%%%%%%%%%%%%%%%
\subsection{Scalar field theory}
\label{app:scalar}
%%%%%%%%%%%%%%%%%%%%%%%%%%%%%%%%%%%%%%%%%%%%%%%%%%%%%%%%%%%%%%%%%%%%%%%%%

In this section we evaluate the necessary phase space integrals for a 
scalar field theory.  Let us first start with the integral labeled 
$\mathcal{F}$ in eq.~\ref{eq:F},
\beqa
\mathcal{F} = \intPSb \left(\frac{p^2}{3}-c_s^2 \Ep 
\frac{\partial \left(\beta\Ep\right) }{\partial \beta}\right)\np(1+\np)\;.
\label{eq:F123}
\eeqa
In the high temperature limit we can take $\beta m\to 0$ while keeping 
$\tilde{m}$ finite.  Using the phase space integrals defined in 
appendix~\ref{app:RTA} we find\footnote{In appendix~\ref{app:RTA} the 
phase space integral in eq.~(\ref{eq:RTA39}) has the form $\int \p^6/
\Ep^{1+\alpha}$.  The term in eq.~\ref{eq:F123} proportional to 
$\tilde{m}^2$ has from $\int \p^{6-\alpha}/\Ep$.  For massless particles 
these two integrals are the same except if $\alpha=4$ in the case of 
bosons.  This is due to the way the logarithmic divergence is regulated 
in the two cases.  In the latter case where there is only one power of 
$\Ep$ in the denominator we define
\beqa
\Gt(x=2)\equiv\lim_{m\to 0}\beta^2\int \frac{\p^2}{\Ep}\np(1\pm\np) dp=
\left\{\begin{array}{l c}
\Gamma(2) & \textrm{Maxwell}\\
\ln\left(\frac{2T}{m}\right) & \textrm{Bose} \\
\Gamma(2)\ln(2) & \textrm{Fermi} \\
\end{array}\right.
\eeqa
}
\beqa
\mathcal{F}=\frac{T^4}{2\pi^2}\left[\left(\frac{1}{3}-c_s^2\right)\G(4)
-(\tilde{m}\beta)^2c_s^2\Gt(2)\right]\, . 
\eeqa
The function $\mathcal{F}$ characterizes the deviation from conformality.  
The relationship between the shifted mass $\tilde{m}$ and the sound speed 
can be found by using the fact that the source term for bulk viscosity is 
orthogonal to the energy--changing zero mode, 
\beqa
0=\intPSa \np(1\pm\np)\left(\frac{\p^2}{3}-c_s^2\tEp^2\right).
\eeqa
This leads to
\beqa
\frac{1}{3}-c_s^2\approx(\tilde{m}\beta)^2\frac{\G(3)}{3\G(5)}\;.
\eeqa
Using this relation we find for bosons
\beqa
\mathcal{F}=\frac{(\tilde{m}T)^2}{6\pi^2}
\left[\frac{15\zeta_+(3)}{2\pi^2}-\ln\left(\frac{2T}{m}\right)\right]\, . 
\eeqa

\end{document}